\documentstyle[a4wide]{article}
\begin{document}

\newcommand{\lm}{\mbox{$\mit\Lambda$}}
\newcommand{\w}{\mbox{$\Omega$}}
\newcommand{\tr}{\mbox{Tr}}
\newcommand{\lsim}{\mbox{\raisebox{-.9ex}{~$\stackrel{\mbox{$<$}}{\sim}$~}}}
\newcommand{\gsim}{\mbox{\raisebox{-.9ex}{~$\stackrel{\mbox{$>$}}{\sim}$~}}}

\begin{titlepage}
\pagestyle{empty}
\begin{small}
\baselineskip=12pt
\rightline{IFIC/01-38}
\rightline{HD-THEP-01-30}
\rightline{Imperial/TP/0-01/23}
\rightline{DAMTP-2001-66}
\rightline{astro-ph/0108093}
\rightline{August 2001}
\end{small}
\baselineskip=21pt

\vspace{1cm}

\begin{center}
     {\Large {\bf NATURAL MAGNETOGENESIS FROM INFLATION }}

\end{center}

\vspace{0.5cm}

\begin{center}
\begin{large}
    {\bf K.~Dimopoulos},$^1$
    {\bf T.~Prokopec},$^2$ 
    {\bf O.~T\"{o}rnkvist}$^3$ {\bf and}
    {\bf A.~C.~Davis}$^4$
\end{large}

\bigskip

\baselineskip=12pt

$^1${\it Instituto de F\'{\i}sica Corpuscular,
Universitat de Valencia/CSIC,}\\
{\it Apartado de Correos 22085, 46071 Valencia, Spain}\\

$^2${\it Institute for Theoretical Physics, Heidelberg University,}\\
{\it Philosophenweg 16, D-69120 Heidelberg, Germany}\\

$^3${\it Theoretical Physics Group, Imperial College}\\
     {\it Prince Consort Road, London SW7 2BZ, U.K.} 

$^4${\it Department of Applied Mathematics and Theoretical Physics,
University of Cambridge,}\\
{\it Wilberforce Road, Cambridge CB3~0WA, United Kingdom}\\

\end{center}

\baselineskip=21pt
\vskip 0.2 in

\centerline{ {\bf Abstract} }
\baselineskip=12pt
\vskip 0.3truecm
\noindent
 
\begin{small}
We consider the gravitational generation of the massive $Z$ boson field of 
the standard model, due to the natural breaking of its conformal invariance 
during inflation. The electroweak symmetry restoration at the end of inflation
turns the almost scale-invariant superhorizon $Z$ spectrum into a 
hypermagnetic field, which transforms into a regular magnetic field at the 
electroweak phase transition. The mechanism is generic and is shown to 
generate a superhorizon spectrum of the form $B\propto 1/\ell$ on a 
length-scale $\ell$ 
regardless of the choice of inflationary model. Scaled to the epoch 
of galaxy formation such a field suffices to trigger the galactic dynamo 
and explain the observed galactic magnetic fields in the case of a spatially 
flat, dark energy dominated Universe with GUT-scale inflation. 
The possibility of further amplification 
of the generated field by preheating is also investigated. To this end we 
study a model of Supersymmetric Hybrid Inflation with a Flipped~SU(5) grand 
unified symmetry group.
\end{small}

\vspace{4cm}

{}~\\[-3mm]
\rule{6cm}{.2mm}\newline
\noindent{$^1${\footnotesize
kostas@flamenco.ific.uv.es},
$^2${\footnotesize T.Prokopec@ThPhys.Uni-Heidelberg.DE}}\\
\noindent{$^3${\footnotesize o.tornkvist@ic.ac.uk},
$^4${\footnotesize acd@damtp.cam.ac.uk}}
\end{titlepage}

\section{Introduction}

Magnetic fields permeate most astrophysical systems \cite{kron} and their 
presence may have numerous cosmological and astrophysical implications. 
Indeed, magnetic fields may substantially influence the formation process 
of large-scale structure \cite{struc} and of individual galaxies 
\cite{gal}\cite{rees}. However, the most important role of large-scale
magnetic fields is that they may be responsible for the magnetic fields in
galaxies.

It is a well-known observational fact that galaxies feature magnetic fields
of strength $\sim\mu$Gauss \cite{kron}\cite{beck}\cite{rep}. The structure of 
such fields in spiral galaxies follows closely the spiral pattern \cite{beck} 
and this strongly suggests that the galactic magnetic fields are generated
and sustained by a dynamo mechanism \cite{kron}\cite{beck}\cite{dynamo}.
According to the galactic dynamo, the cyclonic turbulent motion of ionized gas 
combined with the differential galactic rotation amplifies a weak seed field 
exponentially until the backreaction of the plasma motion counteracts the 
growth of the field and stabilizes it to dynamical equipartition 
strength. However, the origin of the required seed field remains elusive.

In order to trigger successfully the galactic dynamo, the seed field has to
satisfy certain requirements of strength and coherence. Indeed, it has been 
shown that seed fields which are too incoherent may destabilize and destroy 
the dynamo action \cite{kuls}. Most dynamos require a minimum coherence length 
comparable to the dimensions of the largest turbulent eddy, $\sim$~100~pc. 
The required strength is determined by the dynamo's amplification 
timescale (typically the galactic rotation period) and the age of the 
galaxy. Recent observations suggest that the Universe at present is dominated 
by a dark-energy component \cite{lamda}. In such a case 
the galaxies are older 
than previously thought, and the minimum seed-field strength may be as low 
as \mbox{$B_{\rm seed}\sim 10^{-30}$Gauss} \cite{dlt}.

There have been attempts to generate the necessary seed field using various
astrophysical mechanisms, the most important of which involve battery 
\cite{batt} or vorticity effects \cite{rees}\cite{vort}. Battery 
mechanisms require a large-scale misalignment of density and pressure 
(temperature) gradients, usually 
associated with large lobe-jets (AGNs) or starburst activity \cite{batt2},
and are therefore difficult to realize in the majority of the galaxies. 
On the other hand, large-scale vortical motions can be effective 
only if the ionization of the plasma is substantial, which can hardly occur 
as late as the epoch of galaxy formation. 

Due to the above difficulties, it has been frequently argued that the origin 
of the seed field may precede the galaxies themselves, i.e. be truly 
primordial. The obvious requirement for large-scale magnetic-field generation 
before the time of 
recombination is that it has to occur out of thermal equilibrium, because
such a field breaks isotropy \cite{sasha}. This limits the choice to 
generating the magnetic field either 
at a phase transition or during inflation, 
although more exotic mechanisms have occasionally been suggested, such as 
vorticity-inducing cosmic strings \cite{CS} or magnetic fields in the 
pre-Big Bang era \cite{PBB}.

There have been numerous attempts to create a primordial magnetic field at the 
breaking of grand unification or at the electroweak phase transition or even 
at the quark-confinement epoch (QCD transition) \cite{PT}. However, since the 
generating mechanisms are causal, the coherence of the created magnetic field 
cannot be larger than the particle horizon at the time of the phase 
transition. Because all the above transitions occurred very early in the 
Universe's history, the comoving size of the horizon is rather small (the best 
case is the QCD transition, for which the horizon corresponds to $\sim$~1~AU) 
and so the resulting magnetic fields are too incoherent. It has been shown 
that not even the most favorable turbulent evolution can adequately increase 
the correlation length of such fields \cite{kdacd}. Thus, one should achieve 
superhorizon correlations in order to generate a sufficiently coherent 
magnetic field in the early Universe. 

Inflation presents the only known way to achieve correlations beyond the 
horizon scale and, for this reason, has received a lot of attention.
However, the prime obstacle to generating magnetic fields during inflation
is the conformal invariance of electromagnetism, which forces the magnetic 
flux to be conserved \cite{TW}. 
As a result, the strength of any generated magnetic field 
decreases exponentially as the inflationary Universe rapidly expands.
Attempts to overcome this problem have set out to break the conformal 
invariance in various ways, such as through the explicit introduction of 
terms in the Lagrangian 
which couple the photon directly to gravity or to a scalar or pseudoscalar 
field, or through the inclusion of a massive photon, or even 
by means of the conformal anomaly \cite{TW}\cite{rest}. 
However, those very few attempts that succeed in producing
a sufficiently strong magnetic field do so relinquishing simplicity.
Recently, interesting efforts have been made to create magnetic fields by 
coupling the photon to some scalar field during inflation \cite{calz} or at 
the preheating stage \cite{bassett}. 
These proposals have since been criticized in \cite{gs}.
Other recent ideas include magnetogenesis due to the breakdown of Lorentz 
invariance in the context of string theory and non-commutative VSL theories
\cite{vsl}, due to the dynamics of large extra dimensions \cite{extra}
and, finally, due to gauge field coupling to metric perturbations 
\cite{maroto}. The viability of the latter has been questioned in \cite{tom}.

Recently one of us (TP) has studied the effects of conformal symmetry 
breaking due to the coupling of the photon field to fermions and scalars 
\cite{tom}. The author has considered effective actions  
arising from loop corrections in the $1/M$ expansion and from the anomaly.
In addition, the photon coupling to scalar and pseudoscalar fields was 
reconsidered, having been originally introduced by Ratra in \cite{rest} and in 
\cite{TW} (see also Garretson {\em et al.} in \cite{rest} and 
\cite{giova}).

In this paper we show that conformal invariance is naturally broken in
inflation without the need for
 any exotic mechanism or field. In particular, we 
consider the gravitational generation of the massive $Z$-boson field of the 
standard model. The electroweak symmetry restoration at the end of inflation
turns the superhorizon $Z$ spectrum into a hypermagnetic field, which 
transforms into a regular magnetic field at the electroweak phase transition.
The mechanism is rather generic and is shown to generate a superhorizon
spectrum of the form \mbox{$B_{\rm rms}\propto 1/\ell$} on a length-scale 
$\ell$ regardless of the choice of inflationary model. However, it is 
possible to consider further amplification of the generated field {\em via\/} 
parametric resonance during preheating. To this
end, we study a model of Supersymmetric Hybrid Inflation (SUSY-HI)
with Flipped~SU(5)
as the symmetry group of the Grand Unified Theory (GUT). Our mechanism for
creating the field has been described in \cite{letter}, 
where it was not specifically applied to the $Z$ field. 

The structure of our paper is as follows. In Section~2, the
model of Supersymmetric Hybrid Inflation is presented. Section~3 contains a 
study of the relevant particle representations of the
Flipped~SU(5) group. The symmetry breaking process is 
analyzed and the field equations of the gauge bosons are obtained, with 
particular emphasis on the hypercharge 
field and its source current. Also, the 
equations for the scalar fields (Higgs fields and the inflaton) are 
layed out in detail. In Section~4, we study the gravitational production of 
the $Z$-boson field during inflation in a model-independent way. The resulting 
spectrum of the magnetic field is computed and scaled down to the epoch of 
galaxy formation, thus showing that it may be sufficient for explaining the 
galactic magnetic fields. In Section~5, the possibility of extra amplification 
by preheating is investigated using the particular model of Flipped-SU(5)
SUSY-HI. Both Flipped SU(5) and SUSY-HI are well 
motivated by Supergravity and Superstrings and rather are naturally compatible.
However, it should be noted that the choice of GUT 
group was made only to allow us to perform analytical calculations. In 
principle, any GUT group would suffice. Finally, in Section 6 we present our 
conclusions. We have attached two Appendices, one describing the Hartree 
approximation used at certain points in the calculations and the other 
presenting the way to calculate the rms amplitudes of the $Z$-boson and the 
hypermagnetic field.

Throughout the paper we use a $(+,-,-,-)$ metric 
and units with \mbox{$c=\hbar=1$} so that Newton's gravitational constant is 
\mbox{$8\pi G=m_P^{-2}$}, where \mbox{$m_P=2.4\times 10^{18}$GeV} is the 
reduced Planck mass.

\section{Supersymmetric Hybrid Inflation}

\subsection{Hybrid Inflation}

Hybrid Inflation was originally suggested by Linde \cite{linde} to 
avoid the fine-tuning problems of most inflationary models due to radiative 
corrections. In order to achieve this, Hybrid Inflation 
introduces a mass scale $M$, much smaller than the Planck mass, that sets the 
scale of the false-vacuum energy towards the end of inflation. Thus,
at least near the end, the inflationary potential is protected from 
radiative corrections and is sufficiently flat. The expense paid
is the necessity of introducing, in addition to the inflaton field
$s$, another scalar field $\phi$ related to the scale $M$. 
The scalar potential for Hybrid Inflation is, 

\begin{equation}
V(\phi,s)=\frac{1}{4}\lambda
\left(\phi^2-M^2\right)^2+
\frac{1}{2}hs^2\phi^2+V_s(s)
\label{V}
\end{equation}
where $\lambda$ and $h$ are coupling constants and $V_s(s)$ is the 
slow-roll potential for the inflaton field. 
Because of the coupling 
between $s$ and $\phi$, for \mbox{$s\geq s_c$}, where
\mbox{$s_c\equiv M\sqrt{{\lambda}/{h}}$}, 
the above potential has a global minimum at 
\mbox{$\phi=0$}, so that \mbox{$V(0,s)=\lambda M^4/4$} and we 
obtain low energy-scale inflation as required. However, when 
\mbox{$s< s_c$}, spontaneous symmetry breaking displaces the
minimum of the potential (\ref{V}) to
\mbox{$\phi=M_{e\!f\!f}$}, where 
\mbox{$M_{e\!f\!f}^2\equiv M^2[1-(s/s_c)^2]$}. The system rapidly rolls 
towards the new minimum and oscillates around it, thus terminating 
inflation. In general, inflation ends abruptly less than one $e$-folding 
after the symmetry breaking.

Letting $M$ be of the order of the grand-unification scale can indeed
satisfy the requirements of successful inflation, for which the total
number of $e$-foldings must be large enough (typically about 60) to solve the 
flatness and horizon problems and to account for the
magnitude of density perturbations and the anisotropy of the Cosmic Microwave 
Background Radiation (CMBR) in agreement with Large-Scale Structure and COBE 
observations. Thus, a natural candidate for the $\phi$ scalar is the 
Higgs field of a Grand Unified Theory (GUT). Such a choice has the 
merit of not introducing any additional unknown scalars or mass scales. 

However, the most important attribute of Hybrid Inflation is that it can
originate from Supersymmetry and is, therefore, one of the few inflationary 
models with a theoretical foundation in particle physics.

\subsection{Supersymmetric model}

The literature on Supersymmetric Hybrid Inflation (SUSY-HI) is rather rich, as
it is possible to attain from  either Supergravity or Superstrings 
\cite{hybrid}. We shall briefly describe 
an $F$-term GUT inflationary model as in \cite{fterm}. 
$D$-term models also exist \cite{dterm}, but will not concern us here.

The most general renormalizable 
superpotential with $R$-symmetry is \cite{fterm},
\begin{equation}
W=\kappa S(\Phi\bar\Phi-M^2)
\label{sW}
\end{equation}
where $\Phi\bar\Phi$ is a conjugate pair of singlet components of 
chiral superfields that belong to a nontrivial representation of the
GUT group $G$, $S$ is a gauge-singlet superfield, and
$\kappa$, $M$ are constants that can be made positive by phase 
redefinitions. Introducing the above expression 
to the $F$-term scalar potential 
\mbox{$V_F\simeq \sum_i |{\partial W}/{\partial\phi_i}|^2$}
one finds,
\begin{equation}
V=\kappa^2|M^2-\Phi\bar\Phi|^2+\kappa^2|S|^2(|\Phi|^2+|\bar\Phi|^2)
\end{equation}
where we write the scalar components with the same symbols as 
the superfields. Using $R$-symmetry we can bring the scalar fields onto
the real axis. We set \mbox{$S\equiv s/\sqrt{2}$} and 
\mbox{$\Phi=\bar\Phi\equiv\phi$}, where $s$ and $\phi$ are real 
scalar fields.\footnote{The $D$-term vanishes because \mbox{$\Phi=\bar\Phi$}
corresponds to the $D$-flat direction that contains the supersymmetric vacua.}
Then the scalar potential becomes,

\begin{equation}
V(\phi,s)=\kappa^2(\phi^2-M^2)^2+\kappa^2s^2\phi^2
\label{susyV}
\end{equation}
This is, in fact, the potential for Hybrid Inflation given 
in (\ref{V}) with 
\mbox{$\lambda\!=\!2h\!=4\kappa^2$} and 
\mbox{$s_c\equiv \sqrt{2}M$}. 

The non-vanishing vacuum energy breaks supersymmetry and generates 
radiative corrections, which induce the required slow-roll potential 
$V_s(s)$. It can be shown that the
overall contribution is of the form \cite{fterm},
\begin{equation}
V_s(s)\simeq M^4
\frac{\kappa^4}{16\pi^2}(\ln\frac{\sqrt{\kappa}\,s}{\Lambda}+\frac{3}{2})
\label{Vlog}
\end{equation}
where $\Lambda$ is a suitable renormalization mass scale.
Thus, the radiative corrections provide a gentle down-slope for the inflaton,
\mbox{$V_s\propto\ln s$}, which helps to drive it towards its 
minimum. 

At earlier stages of the inflaton's evolution, $s$ may be of the order of 
the Planck mass, so that Supergravity corrections should also be considered.
However, it can be shown \cite{linde2} that the flatness of the potential
is preserved for minimal K\"{a}hler potential (\mbox{$K=2\phi^2+S^2$}).

\section{Flipped SU(5)}

In choosing the GUT for the symmetry breaking that terminates Hybrid 
Inflation, one has to ascertain that there is no monopole problem 
\cite{rachel2}. Thus, a simple or semi-simple group is not an option. We 
decided in favor of Flipped~SU(5)
because of its simplicity and its resemblance to the Standard Model (SM).
Flipped~SU(5) is jargon for the GUT group SU(5)$\times\overline{\mbox{U}}$(1),
in which the hypercharge U(1)$_Y$ is contained both in the SU(5) part and 
the $\overline{\mbox{U}}$(1), in contrast to the Georgi-Glashow 
SU(5)$\times$U(1) model, which has the hypercharge fully embedded in 
SU(5) \cite{flsu5}. 

The supersymmetric version of Flipped~SU(5) is well motivated by 
Superstrings \cite{susyflsu5}. Moreover, it can be considered as an
intermediate stage in the breaking of supersymmetric SO(10) 
\cite{flsu5}\cite{so10}:

\begin{eqnarray}
\mbox{SO(10)} & \rightarrow\;\;\; 
\mbox{SU(5)$\times\overline{\mbox{U}}$(1)}\;\;\; \rightarrow & 
\mbox{SU(3)$_c\times$SU(2)$\times$U(1)$_Y$}\nonumber
\end{eqnarray}

Since the breaking of SO(10) to Flipped~SU(5) generates monopoles, it 
would have to
take place before or during the inflationary period, so that the 
monopoles can be 
safely inflated away.
The SUSY-HI in this model has been studied in \cite{rachel}.

\subsection{The model}

The Lagrangian density of the model is \cite{anne},
\begin{equation}
{\cal L}=-\frac{1}{4}g^{\mu\rho}g^{\nu\sigma}
	\tr(G_{\mu\nu}G_{\rho\sigma})-
	\frac{1}{4}g^{\mu\rho}g^{\nu\sigma}
	G^0_{\mu\nu}G^0_{\rho\sigma}+
	\frac{1}{2}g^{\mu\nu}\tr[(D_\mu\Phi)^\dag D_\nu\Phi]+
	\frac{1}{2}g^{\mu\nu}(\nabla_\mu s)\nabla_\nu s-V\,
\label{L0}
\end{equation}
where $\Phi$ is the GUT-Higgs field, 
$s$ is the inflaton field,
$G_{\mu\nu}$ and $G^0_{\mu\nu}$ are the field strengths of the 
SU(5) and the $\overline{\mbox{U}}$(1) gauge fields respectively, 
$D_\mu\Phi$ is the covariant derivative of the Higgs field, and 
$V$ is the effective potential.

The SU(5) field strength is,
\begin{equation}
G_{\mu\nu}=\nabla_\mu\lm_\nu-\nabla_\nu\lm_\mu-ig[\lm_\mu,\lm_\nu]
\label{H}
\end{equation}
where $g$ is the gauge coupling of SU(5),
\mbox{$\lm_\mu=\lm_\mu^aT^a$}, and $\lm_\mu^a$  ($a=1...24$) 
 are the SU(5) gauge fields with 
$T^a$ being the corresponding generators.
Similarly, for $G^0_{\mu\nu}$ we have,
\begin{equation}
G^0_{\mu\nu}=\nabla_\mu\lm^0_\nu-\nabla_\nu\lm^0_\mu
\label{hatH}
\end{equation}
where $\lm^0_\mu$ is the Abelian gauge field of the $\overline{\mbox{U}}$(1)
group.

Flipped~SU(5) is broken by the GUT Higgs field $\Phi$, which belongs to the 
({\bf 10},1) antisymmetric representation. The $\overline{\mbox{U}}$(1) 
degree of freedom corresponds to an overall phase. 
In this representation, the generators $T^a$ are given by
\mbox{$T^a=-\frac{1}{\sqrt{2}}\mu^a$},
where the set of modified 5$\times$5 Gell-Mann matrices $\mu^a$ can be found 
in the appendix of \cite{anne}. Also, for the Abelian generator we can write 
\mbox{$T^0=-\frac{1}{\sqrt{2}}\mu^0$} and
\mbox{$\mu^0\equiv \sqrt{\frac{3}{5}}\,I$} with $I$ being the 
5$\times$5 identity matrix. The normalization constant has been 
chosen so as to simplify the treatment of the symmetry breaking process.

Writing \mbox{$G_{\mu\nu}=G_{\mu\nu}^aT^a$} and 
\mbox{$\lm_\mu=\lm^a_\mu T^a$}, the field strength (\ref{H}) becomes,
\begin{equation}
G^a_{\mu\nu}=\partial_\mu\lm^a_\nu-\partial_\nu\lm^a_\mu+
gf^{abc}\lm^b_\mu\lm^c_\nu
\label{Ha}
\end{equation}
where we have used that \mbox{$[T^a,T^b]=if^{abc}T^c$} and 
\mbox{$\tr(T^aT^b)=\delta^{ab}$} with $f^{abc}$ being the structure constants 
of the group. Thus, in (\ref{L0}) we have
\mbox{$\tr(G_{\mu\nu}G_{\rho\sigma})=G_{\mu\nu}^aG_{\rho\sigma}^a$}.

The covariant derivative is,

\begin{equation}
D_\mu\Phi=\partial_\mu\Phi-igd(\lm_\mu;\Phi)-
i\bar{g}d(\;\hat{\!\!\lm}_\mu;\Phi)
\label{covd}
\end{equation}
where $\bar{g}$ is the $\overline{\mbox{U}}$(1) gauge coupling,
\mbox{$\;\hat{\!\!\lm}_\mu=\lm^0T^0$}, and,

\begin{equation}
\,[d(\lm_\mu;\Phi)]_{ij}\equiv 
(\lm_\mu)_{ik}\Phi_{kj}+(\lm_\mu)_{jk}\Phi_{ik}
\end{equation}

Since $\Phi_{ij}$ is antisymmetric, the above can be written in matrix 
notation as,

\begin{equation}
d(\lm_\mu;\Phi)=\lm_\mu\Phi-(\lm_\mu\Phi)^T=\lm^a_\mu[T^a\Phi-(T^a\Phi)^T]
\end{equation}
and, similarly, 
\mbox{$d(\;\hat{\!\!\lm}_\mu;\Phi)=-\sqrt{6/5}\;\lm^0_\mu\Phi$}. 

The matrix of the gauge fields, which is hermitian, may be written as,

\begin{equation}
\begin{array}{l}
\w_\mu\equiv -\sqrt{2}\,
\Big(\lm_\mu+\mbox{\Large $\frac{\bar{g}}{g}$}\,\lm^0_\mu\Big)=\\
\\
\left(
\begin{array}{ccccc}
\!\!G^2\!+\!\frac{1}{\sqrt{3}}G^5\!+\!c'_VV\!+\!c_YY\!\!\!\!\!\!\!\!\!\!
 & 
\overline{G}^1 & \overline{G}^3 & \overline{X}^1 & 
\overline{Y}^1 \\
 & & & & \\
G^1 & \!\!\!
\!\!\!\!\!\!\!-G^2\!+\!\frac{1}{\sqrt{3}}G^5\!+\!c'_VV\!+\!c_YY\!\!\!\!\!\!\!
\!\!\!
 & \overline{G}^4 & 
\overline{X}^2 & \overline{Y}^2 \\
 & & & & \\
G^3 & G^4 & \!\!\!
\!\!\!\!\!\!\!-\frac{2}{\sqrt{3}}G^5\!+\!c'_VV\!+\!c_YY&\!\!\!\!\!\!\! 
\overline{X}^3 & 
\overline{Y}^3 \\
 & & & & \\
X^1 & X^2 & X^3 & \!\!\!\!\!\!\!W^3\!-\!\frac{1}{c_V}V\!\! & W^+ \\
 & & & & \\
Y^1 & Y^2 & Y^3 & W^- & \!\!-W^3\!-\!\frac{1}{c_V}V\!\!
\end{array}
\right)_{\mbox{\small $\!\mu$}}
\end{array}
\label{W}
\end{equation}
where we have suppressed the spacetime indices inside the 
matrix. In the above we have
employed the following definitions:~\footnote{In what follows in this Section 
the over-bar denotes charge conjugation 
[e.g. \mbox{$\overline{X}^\alpha_\mu\equiv(X^\alpha_\mu)^*$}]
except in $\bar{g}$ and $\overline{\mbox{U}}$(1).}

\begin{equation}
\begin{array}{lll}
X^1_\mu=\lm^9_\mu+i\lm^{10}_\mu & \hspace{2cm} & 
G^1_\mu=\lm^1_\mu+i\lm^2_\mu \\
 & & \\
X^2_\mu=\lm^{11}_\mu+i\lm^{12}_\mu & & G^2_\mu=\lm^3_\mu \\
 & & \\
X^3_\mu=\lm^{13}_\mu+i\lm^{14}_\mu &  & G^3_\mu=\lm^4_\mu+i\lm^5_\mu \\
 & & \\
Y^1_\mu=\lm^{16}_\mu+i\lm^{17}_\mu & & G^4_\mu=\lm^6_\mu+i\lm^7_\mu \\
 & & \\
Y^2_\mu=\lm^{18}_\mu+i\lm^{19}_\mu & & G^5_\mu=\lm^8_\mu \\
 & & \\
Y^3_\mu=\lm^{20}_\mu+i\lm^{21}_\mu & & W^+_\mu=\overline{W^-_\mu} \\
 & & \\
W^-_\mu=\lm^{22}_\mu+i\lm^{23}_\mu & & W^3_\mu=\lm^{24}_\mu \\
\end{array}
\label{XYG}
\end{equation}
where

\begin{equation}
c_Y = \sqrt{\frac{5}{3}}\,\frac{g_Y}{g} \hspace{1cm}
c_V = \sqrt{\frac{5}{3}}\,\frac{g_Y}{\bar{g}} \hspace{1cm}
c'_V = c_V-\frac{1}{c_V} \hspace{1cm}
g_Y^2=\frac{g^2\bar{g}^2}{g^2+\bar{g}^2}
\label{c}
\end{equation}
and also,

\begin{equation}
V_\mu=
\sqrt{\frac{3}{5}}\,c_V
\Big(\lm^{15}_\mu-\frac{\bar{g}}{g}\,\lm^0_\mu\Big)
\hspace{2cm}
Y_\mu=
\sqrt{\frac{3}{5}}\,c_Y
\Big(\lm^{15}_\mu+\frac{g}{\bar{g}}\,\lm^0_\mu\Big)
\label{VY}
\end{equation}

In the above, the complex bosons $X^\alpha_\mu,Y^\alpha_\mu$ 
($\alpha=1,2,3$) are the supermassive GUT-bosons, $G^i_\mu$ are the 
massless SU(3)$_{\rm c}$ bosons, $W^\pm_\mu$ and $W^3_\mu$ are the usual
SM $W$-bosons and $Y_\mu$ is the hypercharge gauge boson, 
which should not be
confused with the massive $Y^\alpha_\mu$'s. Finally,
$V_\mu$ is a massive boson generated partly by SU(5) and partly by 
$\overline{\mbox{U}}$(1). It can be viewed as the analogue of the $Z$-boson
of the SM. 

In analogy with the $\w_\mu$ matrix, one can define the matrix

\begin{equation}
H_{\mu\nu}\equiv -\sqrt{2}\,
\Big(G_{\mu\nu}+\frac{\bar{g}}{g}\,G^0_{\mu\nu}\Big)
\end{equation}

Using this and (\ref{H}) it is easy to show that

\begin{equation}
H_{\mu\nu}=\partial_\mu\Omega_\nu-\partial_\nu\Omega_\mu+
\frac{ig}{\sqrt{2}}[\Omega_\mu,\Omega_\nu]
\end{equation}

Without loss of generality the GUT Higgs field may be chosen to lie in the 
following direction:

\begin{equation}
\Phi=\frac{\phi}{\sqrt{2}}
\left(
\begin{array}{ccccc}
0 & 0 & 0 & 0 & 0 \\
0 & 0 & 0 & 0 & 0 \\
0 & 0 & 0 & 0 & 0 \\
0 & 0 & 0 & 0 & 1 \\
0 & 0 & 0 & -1 & 0
\end{array}
\right)\quad
\label{phi}
\end{equation}
where $\phi$ is a real positive function of time \mbox{$\phi=\phi(t)$}.
In this case, it can be shown that 
the interaction term of the Lagrangian density becomes,

\begin{equation}
{\cal L}_{\rm int}\equiv
\frac{1}{2}g^{\mu\nu}\tr[(D_\mu\Phi)^\dag D_\nu\Phi]=
\frac{1}{2}g^{\mu\nu}(\partial_\mu\phi)(\partial_\nu\phi)+
\frac{1}{2}M_{XY}^2g^{\mu\nu}
(X_\mu^\alpha\overline{X}_\nu^\alpha+Y_\mu^\alpha\overline{Y}_\nu^\alpha)+
\frac{1}{2}M_V^2g^{\mu\nu}V_\mu V_\nu
\end{equation}
where the masses of the supermassive GUT-bosons and of $V_\mu$ are

\begin{equation}
M_{XY} = \frac{1}{\sqrt{2}}g\phi\hspace{1cm}
M_V=\frac{2}{c_V}M_{XY}
\label{masses}
\end{equation}

Thus, the interaction term reduces to the mass terms of the massive
bosons plus the kinetic term of $\phi$. Note that after the GUT symmetry 
breaking the $W$'s and the hypercharge $Y_\mu$ remain massless. 

It is straightforward to show that the kinetic term of the supermassive 
GUT bosons is,

\begin{equation}
{\cal L}_{\rm kin}^{XY}=
-\frac{1}{4}g^{\mu\rho}g^{\nu\sigma}
H^{X^\alpha}_{\mu\nu}\overline{H}^{X^\alpha}_{\rho\sigma}
-\frac{1}{4}g^{\mu\rho}g^{\nu\sigma}
H^{Y^\alpha}_{\mu\nu}\overline{H}^{Y^\alpha}_{\rho\sigma}
\end{equation}
where\footnote{Because the $H_{\mu\nu}$
matrix is also hermitian, we have:
\mbox{$\tr(H_{\mu\nu}H_{\rho\sigma})=
(H_{\mu\nu})_{ij}(\overline{H}_{\rho\sigma})_{ij}$}.}
\mbox{$H^{X^\alpha}_{\mu\nu}\equiv (H_{\mu\nu})_{4\alpha}$}
\ and \ \mbox{$H^{Y^\alpha}_{\mu\nu}\equiv (H_{\mu\nu})_{5\alpha}$} 
with \ \mbox{$\alpha=1,2,3$}.

\subsection{The field equations of the gauge bosons}

The field equations of the massive GUT-bosons are,

\begin{eqnarray}
\![\partial_\mu+(\partial_\mu\ln\sqrt{-D_g})]
\Big[g^{\mu\rho}g^{\nu\sigma}
(\partial_\rho X_\sigma^\alpha-\partial_\sigma X_\rho^\alpha)\Big]
+M_{XY}^2\,g^{\mu\nu}X_\mu^\alpha
 & = & J^\nu_{4\alpha}\label{feX}\\
 & & \nonumber\\
\![\partial_\mu+(\partial_\mu\ln\sqrt{-D_g})]
\Big[g^{\mu\rho}g^{\nu\sigma}
(\partial_\rho Y_\sigma^\alpha-\partial_\sigma Y_\rho^\alpha)\Big]
+M_{XY}^2\,g^{\mu\nu}Y_\mu^\alpha
 & = & J^\nu_{5\alpha}\label{feYa}
\end{eqnarray}
where \mbox{$D_g\equiv$ det($g_{\mu\nu}$)} and the current, in matrix 
notation, is,\footnote{Note that the current matrix is also hermitian, 
\mbox{$J^\nu_{ij}=\overline{J}^\nu_{ji}$}.}

\begin{equation}
J^\nu=-\frac{ig}{\sqrt{2}}
\Big\{[\partial_\mu+(\partial_\mu\ln\sqrt{-D_g})]
g^{\mu\rho}g^{\nu\sigma}
[\w_\rho,\w_\sigma]+
g^{\mu\rho}g^{\nu\sigma}
[\w_\mu,H_{\rho\sigma}]\Big\}
\label{J}
\end{equation}

The field equations for the massless $W$-bosons are,

\begin{equation}
\,[\partial_\mu+(\partial_\mu\ln\sqrt{-D_g})]
\Big[g^{\mu\rho}g^{\nu\sigma}
(\partial_\rho W^+_\sigma-\partial_\sigma W^+_\rho)\Big]
=J^\nu_{45}
\label{W+0}
\end{equation}

and

\begin{equation}
\,[\partial_\mu+(\partial_\mu\ln\sqrt{-D_g})]
\Big[g^{\mu\rho}g^{\nu\sigma}
(\partial_\rho W^3_\sigma-\partial_\sigma W^3_\rho)\Big]
=\frac{1}{2}(J^\nu_{44}-J^\nu_{55})
\end{equation}

The field equation for $W^-$ is just the complex conjugate of (\ref{W+0}).

The field equations for $Y_\mu$ and $V_\mu$ are,

\begin{eqnarray}
\![\partial_\mu+(\partial_\mu\ln\sqrt{-D_g})]
[g^{\mu\rho}g^{\nu\sigma}(\partial_\rho V_\sigma-\partial_\sigma V_\rho)]
+M_V^2\,g^{\mu\nu}V_\mu
 & = & \mbox{\small $-\sqrt{\frac{5}{12}}$}
\cos\Theta\,J_Y^\nu\label{feV}\\
 & & \nonumber\\ 
\![\partial_\mu+(\partial_\mu\ln\sqrt{-D_g})]
[g^{\mu\rho}g^{\nu\sigma}(\partial_\rho Y_\sigma-\partial_\sigma Y_\rho)]
 & = &\!\! \mbox{\small $-\sqrt{\frac{5}{12}}$}
\sin\Theta\,J_Y^\nu\label{feY}
\end{eqnarray}
where \mbox{$\tan\Theta\equiv\bar{g}/g$} is the GUT equivalent of the 
Weinberg angle and the hypercharge source current is, 

\begin{equation}
J_Y^\nu=J_{44}^\nu+J_{55}^\nu
\label{JY}
\end{equation}

\subsection{More on the hypercharge source current}

It can be shown that the hypercharge source current can be written as,

\begin{equation}
J_Y^\nu=\sqrt{2}\,g\,\mbox{Im}
\Big\{[\partial_\mu+(\partial_\mu\ln\sqrt{-D_g})]
g^{\mu\rho}g^{\nu\sigma}
(X^\alpha_\rho\overline{X}^\alpha_\sigma+
Y^\alpha_\rho\overline{Y}^\alpha_\sigma)
+g^{\mu\rho}g^{\nu\sigma}
\Big(X^\alpha_\mu
\overline{H}^{X^\alpha}_{\rho\sigma}+
Y^\alpha_\mu 
\overline{H}^{Y^\alpha}_{\rho\sigma}\Big)\Big\}
\end{equation}

The above gives,

\begin{eqnarray}
J_Y^\nu & = & \sqrt{2}\,g\,
\mbox{Im}\Big\{[\partial_\mu+(\partial_\mu\ln\sqrt{-D_g})]
g^{\mu\rho}g^{\nu\sigma}
(X^\alpha_\rho\overline{X}^\alpha_\sigma+
Y^\alpha_\rho\overline{Y}^\alpha_\sigma)+\nonumber\\
 & & \nonumber\\
 & & +\,g^{\mu\rho}g^{\nu\sigma}
[X^\alpha_\mu\partial_\rho\overline{X}^\alpha_\sigma
+Y^\alpha_\mu\partial_\rho\overline{Y}^\alpha_\sigma
-(\rho\leftrightarrow\sigma)]\Big\}+\label{cubic}\\
 & & \nonumber\\
 & & +\,g^2g^{\mu\rho}g^{\nu\sigma}
\mbox{Re}[(\w_\mu)_{4k}(\w_\rho)_{kl}(\w_\sigma)_{l4}
+(\w_\mu)_{5k}(\w_\rho)_{kl}(\w_\sigma)_{l5}-
(\rho\leftrightarrow\sigma)]\nonumber
\end{eqnarray}

It is now evident that, only through the last term in the above expression, 
may the hypercharge source current receive contributions 
from gauge fields other than the supermassive GUT-bosons.
It can be shown that the contributions to $J_Y^\nu$ from the various gauge 
fields are of the following form,

The contribution from the hypercharge field:

\begin{equation}
J_Y^\nu[Y]=g^2c_Yg^{\mu\rho}g^{\nu\sigma}\Big[\,Y_\rho\,
\mbox{Re}(X^\alpha_\mu\overline{X}^\alpha_\sigma+Y^\alpha_\mu
\overline{Y}^\alpha_\sigma)-(\rho\leftrightarrow\sigma)\,\Big]
\label{JYY}
\end{equation}

The contribution from massive $V$:

\begin{equation}
J_Y^\nu[V]=g^2c_Vg^{\mu\rho}g^{\nu\sigma}\Big[\,V_\rho\,
\mbox{Re}(X^\alpha_\mu\overline{X}^\alpha_\sigma+Y^\alpha_\mu
\overline{Y}^\alpha_\sigma)-(\rho\leftrightarrow\sigma)\,\Big]
\label{JYV}
\end{equation}

The contribution from $W^3$:

\begin{equation}
J_Y^\nu[W^3]=g^2g^{\mu\rho}g^{\nu\sigma}\Big[\,W^3_\rho\,
\mbox{Re}(Y^\alpha_\mu\overline{Y}^\alpha_\sigma-
X^\alpha_\mu\overline{X}^\alpha_\sigma)
-(\rho\leftrightarrow\sigma)\,\Big]
\label{JYW3}
\end{equation}

The contribution from $W^\pm$:

\begin{equation}
J_Y^\nu[W^\pm]=-g^2g^{\mu\rho}g^{\nu\sigma}
\mbox{Re}\Big[\,
Y^\alpha_\mu\overline{X}^\alpha_\sigma W^{\mbox{\small +}}_\rho+
X^\alpha_\mu\overline{Y}^\alpha_\sigma W^-_\rho
-(\rho\leftrightarrow\sigma)\,\Big]
\label{JYWpm}
\end{equation}

The contribution from the gluons $G^i$:

\begin{equation}
J_Y^\nu[G]= g^2g^{\mu\rho}g^{\nu\sigma}
\mbox{Re}\Big[\,
(X^\alpha_\mu\overline{X}^{_{\mbox{\scriptsize $\beta$}}}_\sigma+
Y^\alpha_\mu\overline{Y}^{_{\mbox{\scriptsize $\beta$}}}_\sigma) 
\w_\rho^{\alpha\beta}-(\rho\leftrightarrow\sigma)\,\Big]
\label{JYG}
\end{equation}
where \mbox{$\alpha,\beta=1,2,3$}. For \mbox{$\alpha=\beta$} one should
consider only the $G$-part of $\w_\rho^{\alpha\beta}$.

In view of all the above we can write the hypercharge field equation as,

\begin{equation}
\begin{array}{l}
\![\partial_\mu+(\partial_\mu\ln\sqrt{-D_g})]
[g^{\mu\rho}g^{\nu\sigma}(\partial_\rho Y_\sigma-\partial_\sigma Y_\rho)]
+\frac{5}{6}g^2_Y
{\cal E}^{\mu\nu\rho\sigma}
\mbox{Re}(X^\alpha_\mu\overline{X}^\alpha_\sigma+Y^\alpha_\mu
\overline{Y}^\alpha_\sigma)\,Y_\rho=\\
\\
=-\frac{5}{6}\cot\Theta\,g_Y^2
{\cal E}^{\mu\nu\rho\sigma}
\mbox{Re}(X^\alpha_\mu\overline{X}^\alpha_\sigma+Y^\alpha_\mu
\overline{Y}^\alpha_\sigma)\,V_\rho-\\
\\
-\sqrt{\frac{5}{6}}
\,g_Y
\mbox{Im}\Big\{[\partial_\mu+(\partial_\mu\ln\sqrt{-D_g})]
g^{\mu\rho}g^{\nu\sigma}
(X^\alpha_\rho\overline{X}^\alpha_\sigma+
Y^\alpha_\rho\overline{Y}^\alpha_\sigma)\,+\\
\\
\hspace{3cm}+\,
g^{\mu\rho}g^{\nu\sigma}
[X^\alpha_\mu\partial_\rho\overline{X}^\alpha_\sigma
+Y^\alpha_\mu\partial_\rho\overline{Y}^\alpha_\sigma
-(\rho\leftrightarrow\sigma)]\Big\}\;-\\
\\
-\sqrt{\frac{5}{12}}
g_Yg\,
{\cal E}^{\mu\nu\rho\sigma}
\Big[W^3_\rho\,
\mbox{Re}(Y^\alpha_\mu\overline{Y}^\alpha_\sigma\!\!-\!
X^\alpha_\mu\overline{X}^\alpha_\sigma)\!-\!
\mbox{Re}(Y^\alpha_\mu\overline{X}^\alpha_\sigma W^{\mbox{\small +}}_\rho
\!\!+\!X^\alpha_\mu\overline{Y}^\alpha_\sigma W^-_\rho)\Big]
\!-\!\!\sqrt{\frac{5}{12}}
\sin\Theta J_Y^\nu\![G]
\end{array}
\label{JYbig}
\end{equation}
where we have used that,

\begin{equation}
g^{\mu\rho}g^{\nu\sigma}
\Big[A_\sigma B_\rho-(\rho\leftrightarrow\sigma)\Big]=
{\cal E}^{\mu\nu\rho\sigma}
A_\sigma B_\rho
\hspace{1cm}
{\cal E}^{\mu\nu\rho\sigma}\equiv
\epsilon_{\kappa\lambda\xi}\epsilon^{\kappa\mu\nu}
g^{\rho\lambda}g^{\sigma\xi}
\label{BigE}
\end{equation}
for any $A_\sigma$ and $B_\rho$, where 
$\epsilon^{\kappa\mu\nu}$ is the totally antisymmetric 
Levi-Civita tensor.

Although the above equation appears rather complicated it can be understood 
as follows. If we ignore the expansion of the Universe then (\ref{JYbig}) 
becomes, schematically,

\begin{equation}
\Box Y+\bar{g}^2_Y(XX)Y=
-\cot\Theta\,\bar{g}_Y^2(XX)V
-\bar{g}_Y\{(X\partial X)+(g/\!\mbox{\footnotesize $\sqrt{2}$})[(XXW)+(XXG)]\}
\label{JYH}
\end{equation}
where we have symbolized all the supermassive GUT-bosons with $X$, we have 
set\footnote{Corresponding to the definition used in \cite{flsu5}.} 
\mbox{$\bar{g}_Y\equiv\sqrt{5/6}\,g_Y$} and
we have taken the equivalent of the Lorentz gauge for the hypercharge field.

\subsection{The field equations for the scalar fields}

The effective potential is taken to be,

\begin{equation}
V=\frac{1}{4}\lambda\left[\,\tr(\Phi^\dag\Phi)-M^2\right]^2+
\frac{1}{2}hs^2\tr(\Phi^\dag\Phi)+V_s(s) 
\label{V0}
\end{equation}
where $M$ is the scale of the GUT symmetry 
breaking, $\lambda$ and $h$ are coupling constants and $V_s(s)$ is the 
slow-roll potential for the inflaton field. 
For the particular $\Phi$-gauge 
introduced in (\ref{phi}) the scalar potential reduces to the one
given in (\ref{V}), for which in SUSY-HI we have 
\mbox{$\lambda=2h\sim 1$} and \mbox{$V_s\propto\ln s$}.

The field equation of the inflaton is,

\begin{equation}
\![\partial_\mu+(\partial_\mu\ln\sqrt{-D_g})]\,g^{\mu\nu}\partial_\nu s=
-\frac{\partial V}{\partial s}
\end{equation}

For a spatially-flat Friedman-Robertson-Walker (FRW) 
metric and with \mbox{$s=s(t)$} the above reduces to 
the well known form,

\begin{equation}
\ddot{s}+3H\dot{s}+\frac{\partial V}{\partial s}=0
\label{inflaton}
\end{equation}
where \mbox{$H=\dot{a}/a$} is the Hubble parameter with \mbox{$a=a(t)$} 
being the scale factor of the Universe and the dot denotes derivative with 
respect to the cosmic time $t$. For the potential (\ref{V}) we have,

\begin{equation}
\frac{\partial V}{\partial s}=h\phi^2s+\frac{\partial V_s}{\partial s}
\label{inflaton1}
\end{equation}

Using a Conformal-time FRW (CFRW) metric and with \mbox{$V_s\propto\ln s$}
we find,

\begin{equation}
\sigma''-\frac{a''}{a}\sigma+(h\varphi^2\sigma+
a^4\frac{\partial V_\sigma}{\partial\sigma})=0
\label{cfrw-s}
\end{equation}
where \mbox{$V_\sigma\propto\ln\sigma$},
the primes denote derivatives with respect to conformal time $\tau$ and,

\begin{equation}
\sigma\equiv a\,s\hspace{2cm}\varphi\equiv a\,\phi
\label{cfrw-sf}
\end{equation}

The field equation for $\Phi$ is more complicated. The general expression
is found to be,

\begin{equation}
\![\partial_\mu+(\partial_\mu\ln\sqrt{-D_g})]\,g^{\mu\nu}(D_\nu\Phi)=
igg^{\mu\nu}\lm^a_\mu d(T^a;D_\nu\Phi)
-\Big\{\lambda\left[\tr(\Phi^\dag\Phi)-M^2\right]+hs^2\Big\}\Phi
\label{higgs}
\end{equation}

The above is a complex matrix equation. The resulting relevant equation for 
$\varphi$ in CFRW is,

\begin{equation}
\varphi''-\frac{a''}{a}\varphi+[\lambda(\varphi^2-a^2M^2)+h\sigma^2]
\,\varphi=
\frac{1}{2}a^2g^2g^{\mu\nu}(X_\mu^\alpha\overline{X}_\nu^\alpha+
Y_\mu^\alpha\overline{Y}_\nu^\alpha
+\frac{4}{c_V^2}V_\mu V_\nu)\,\varphi
\label{cfrw-f}
\end{equation}
where we ignore all but the spatially uniform mode of $\varphi$. 

\subsection{The contribution of the electroweak Higgs field}

Due to supercooling (\mbox{$T\propto a^{-1}\rightarrow 0$}) the 
electroweak (EW) symmetry is broken during inflation and, thus, it is 
important to determine the contribution of the EW-Higgs field $\Psi$
to the field equations of the gauge fields. To that end we have to consider
the embedding pattern of the SM group into our GUT.

The SM group is not fully contained in the SU(5) part of Flipped~SU(5). 
The SU(2) and the SU(3)$_c$ of the SM are contained in SU(5) so that their 
couplings, say $g_2$ and $g_3$ merge and equal $g$ at the GUT scale. But the 
hypercharge coupling $g_Y$ does not. In fact this coupling can be considered 
to be comprised of the $\bar{g}$ coupling of $\overline{U}$(1) and a $g_1$ 
coupling corresponding to a U(1)$_1$ subgroup of the SU(5) \cite{flsu5}. 
This $g_1$ merges with $g_2$ and $g_3$ at the GUT
scale. Thus the structure of the symmetry breaking is,

\bigskip

\begin{tabular}{lcc}
\underline{Flipped-SU(5)} & &  \underline{Standard Model}\\
 & & \\
$\left.
\mbox{\begin{tabular}{lc}
\begin{tabular}{ccc}
SU(5) & $\rightarrow$ & SU(3)$\times$SU(2)\\
 & & $\times$
\end{tabular} & \\
$\left.
\mbox{\begin{tabular}{ccc}
$\times$ & & \hspace{0.5cm}U(1)$_1$ \\                       
 & & \\                                        
$\overline{U}$(1) & $\;\;\rightarrow\;\;$ & \hspace{0.5cm}$\overline{U}$(1)    
\end{tabular}}
\right\}$ & $\!\!\!\!\rightarrow$ U(1)$_Y$ 
\end{tabular}}
\right\}$ & $\rightarrow$ & SU(3)$\times$SU(2)$\times$U(1)$_Y$
\end{tabular}

\bigskip

where

\bigskip

\begin{eqnarray}
\left.
\begin{array}{c}
g_1\\
g_2\\
g_3
\end{array}
\right\}\rightarrow g & \;\mbox{at GUT scale} 
\hspace{1cm}\mbox{and}\; &
g_Y =\frac{g\bar{g}}{\sqrt{g_1^2+\bar{g}^2}}\nonumber
\end{eqnarray}

\bigskip

It is obvious that at the GUT scale, because \mbox{$g_1\rightarrow g$}, 
the hypercharge coupling is given by (\ref{c}) as required.
The U(1)$_1$ generator corresponds to the $T^{15}$ generator of 
SU(5) since this is the one that gets mixed with the generator of 
$\overline{U}$(1) to give the hypercharge generator.

In our framework we have the peculiar situation that, although the GUT
symmetry is unbroken during inflation due to the coupling between the 
inflaton and the GUT-Higgs field, the EW symmetry is, in principle, broken 
because the Universe is supercooled.\footnote{In fact, the quantum 
fluctuations restore also the electroweak symmetry in a sense. What actually
happens is that the EW-Higgs field forms a non-zero condensate, as we will 
explain in Sec. \ref{43}, which provides masses for the massive EW gauge 
bosons, thereby breaking their conformal invariance.}
Therefore, we are interested in finding the contributions of 
the EW-Higgs are to the field equations.

The additional EW contribution to the Lagrangian (\ref{L0}) is of the form,

\begin{equation}
{\cal L}_{EW}=\frac{1}{2}g^{\mu\nu}(D_\mu\Psi)^\dag D_\nu\Psi-
U(\Psi^\dag\Psi)
\label{LEW}
\end{equation}
where,

\begin{equation}
U(\Psi^\dag\Psi)=\frac{1}{4}\lambda_*(\Psi^\dag\Psi-M_{EW}^2)^2
\label{u}
\end{equation}
with $M_{EW}$ being the electroweak scale and

\begin{equation}
D_\mu\Psi=(\nabla_\mu+ig_2W^\alpha_\mu\tau^\alpha+ig_YY_\mu)\Psi
\end{equation}
where $\tau^\alpha$ are the generators of the SU(2) of the SM 
such that \mbox{$W_\mu^\alpha\tau^\alpha=\sqrt{2}\,
\lm_\mu^aT^a$} with \mbox{$a=21+\alpha$}, 
\mbox{$\alpha=1,2,3$} and at GUT scale 
\mbox{$g_2\rightarrow g$}. 

The EW-Higgs field is in the {\bf 5} complex vector representation. 
Because the gauge condition (\ref{phi}) 
leaves the full EW symmetry group unbroken,
we can, without loss of generality, introduce the gauge, 
\mbox{$\Psi^T=\psi$(0 0 0 1 0)}, where $\psi=\psi(t)$ is a real, positive 
function. Then the complete field equations of the $W$'s are found to be,

\begin{equation}
\,[\partial_\mu+(\partial_\mu\ln\sqrt{-D_g})]
\Big[g^{\mu\rho}g^{\nu\sigma}
(\partial_\rho W^+_\sigma-\partial_\sigma W^+_\rho)\Big]
+M_W^2g^{\mu\nu}W^+_\mu
=J^\nu_{45}
\label{W+}
\end{equation}

and 

\begin{equation}
\,[\partial_\mu+(\partial_\mu\ln\sqrt{-D_g})]
\Big[g^{\mu\rho}g^{\nu\sigma}
(\partial_\rho W^3_\sigma-\partial_\sigma W^3_\rho)\Big]
+M_W^2g^{\mu\nu}W^3_\mu
=\frac{1}{2}(J^\nu_{44}-J^\nu_{55})
+\sin\Theta (g\psi)^2g^{\mu\nu}Y_\mu
\end{equation}
where \mbox{$M_W\equiv g\psi$} and the field equation of $W^-_\mu$ is 
just the complex conjugate  of (\ref{W+}).

Moreover for the hypercharge we have,

\begin{equation}
\,[\partial_\mu+(\partial_\mu\ln\sqrt{-D_g})]
[g^{\mu\rho}g^{\nu\sigma}(\partial_\rho Y_\sigma-\partial_\sigma Y_\rho)]
+M_Y^2g^{\mu\nu}Y_\mu
=-\sqrt{\mbox{\normalsize $\frac{5}{12}$}}\sin\Theta\,J_Y^\nu
+\sin\Theta (g\psi)^2g^{\mu\nu}W^3_\mu
\label{newYfe}
\end{equation}
where \mbox{$M_Y\equiv g_Y\psi$}.
It can be shown that the contribution of the EW-Higgs to
the field equation of the massive boson $V_\mu$ is zero, as expected.
Now, let us rotate the above to form the $Z_\mu$ boson and the photon 
$A_\mu$ of the SM. They are defined as,

\begin{eqnarray}
Z_\mu & = & \cos\theta_WW^3_\mu-\sin\theta_WY_\mu\label{ZWY}\\
 & & \nonumber\\
A_\mu & = & \sin\theta_WW^3_\mu+\cos\theta_WY_\mu\label{AWY}
\end{eqnarray}
where $\theta_W$ is the Weinberg angle defined as,
\mbox{$\tan\theta_W\equiv g_Y/g_2$}. Note that at GUT scale
\mbox{$\tan\theta_W=\sin\Theta$}.

Thus, we obtain,

\begin{equation}
\begin{array}{r}
\,[\partial_\mu+(\partial_\mu\ln\sqrt{-D_g})]
[g^{\mu\rho}g^{\nu\sigma}(\partial_\rho Z_\sigma-\partial_\sigma Z_\rho)]
+M_Z^2g^{\mu\nu}Z_\mu
=\\
\\
=\frac{1}{2}\cos\theta_W\Big[
(1+\sqrt{\frac{5}{3}}\sin^2\Theta)J_{44}^\nu-
(1-\sqrt{\frac{5}{3}}\sin^2\Theta)J_{55}^\nu\Big]
\end{array}
\label{Z0}
\end{equation}
where 

\begin{equation}
M_Z\equiv g_Z\psi\hspace{1cm}\mbox{and}\hspace{1cm}
g_Z\equiv \frac{g}{\cos\theta_W} 
\label{MZ}
\end{equation}

Thus, we see that \mbox{$M_Z=M_W/\cos\theta_W$} as expected.
For the photon we find,

\begin{equation}
\,[\partial_\mu+(\partial_\mu\ln\sqrt{-D_g})]
[g^{\mu\rho}g^{\nu\sigma}(\partial_\rho A_\sigma-\partial_\sigma A_\rho)]
=\mbox{\normalsize $\frac{1}{2}$}\sin\theta_W\Big[
(1-\sqrt{\mbox{\normalsize $\frac{5}{3}$}}\,)J_{44}^\nu-
(1+\sqrt{\mbox{\normalsize $\frac{5}{3}$}}\,)J_{55}^\nu\Big]
\label{A0}
\end{equation}

Finally, the field equation of $\Psi$ is,

\begin{equation}
\,[\partial_\mu+(\partial_\mu\ln\sqrt{-D_g})]\,
g^{\mu\nu}(D_\nu\Psi)=
-ig^{\mu\nu}(gW_\mu^\alpha\tau^\alpha+g_YY_\mu)D_\nu\Psi-
\lambda_*(\Psi^\dag\Psi-M_{EW}^2)\Psi
\end{equation}
which, in the gauge mentioned above, becomes,

\begin{equation}
y''-\frac{a''}{a}y+[\lambda_*(y^2-a^2M_{EW}^2)]y=
a^2g^{\mu\nu}
[\,g^2(W_\mu^1W_\nu^1+W_\mu^2W_\nu^2)+g_Z^2Z_\mu Z_\nu]\,y
\label{yfe}
\end{equation}
where

\begin{equation}
y\equiv a\psi
\label{y}
\end{equation}

It should be pointed out here that the gauge boson masses appearing in the
corresponding
field equations above are a purely mathematical result of the structure of
the Flipped~SU(5) group. They will be non-zero only if a Higgs-field 
condensate is generated, such as happens in the relevant phase transition.

\section{Conformal invariance breakdown during inflation}\label{mech}

In this section we will show that the $Z$-field is naturally produced 
during inflation because its mass term is non-zero
and, therefore, the field is not conformally invariant, but instead 
it has gravitational source terms. At the end of inflation reheating 
restores the EW-symmetry and the generated $Z$-spectrum is projected onto
the direction of the massless, Abelian hypercharge field, thus, creating a 
hypermagnetic field, 
which freezes into the primordial plasma and survives until the EW 
phase transition. At that time the hypercharge configuration is 
projected onto the photon, transforming the hypermagnetic field 
into a regular magnetic field, which evolves until galaxy formation. 
We will show that such a field may be strong enough to seed the dynamo in 
galaxies and account for their observed magnetic fields.

\subsection{\boldmath The mode equations for $Z$}\label{modequsZ}

During inflation the temperature is essentially zero, which means that 
the electroweak symmetry is broken. Thus, we expect \mbox{$\psi\neq 0$},
which, in view of (\ref{MZ}) renders the $Z_\mu$ gauge field massive. 
On the other hand, due to the interaction between the inflaton and the 
GUT-Higgs field, the GUT symmetry is unbroken and \mbox{$\phi=0$}. This means 
that the GUT-bosons $X^\alpha_\mu$ and $Y^\alpha_\mu$ are massless. The same 
is true also for the $G^i_\mu$ bosons because they do not couple to any scalar 
field. Therefore, the $X^\alpha_\mu,Y^\alpha_\mu$ and the $G^i_\mu$
bosons remain conformally invariant during inflation so, like the photon,
they cannot be generated gravitationally. Thus, their 
magnitude during inflation is negligible. For this reason, because the 
current in (\ref{Z0}) is primarily sourced by the GUT-bosons we 
will ignore its contribution during inflation. 

Assuming a CFRW metric 
\mbox{$g_{\mu\nu}=a(\tau)^2\eta_{\mu\nu}$} the field equation (\ref{Z0}) 
may be rewritten as,

\begin{equation}
\eta^{\mu\rho}\eta^{\nu\sigma}
\partial_\mu(\partial_\rho Z_\sigma-\partial_\sigma Z_\rho)
+a^2M_Z^2\eta^{\mu\nu}Z_\mu=0
\label{CFRWZ}
\end{equation}
where $\eta^{\mu\nu}$ is the flat spacetime Minkowski metric and
\mbox{$M_Z=M_Z(\tau)$} is the mass of the $Z$-boson generated
primarily by the self-interaction term of the EW Higgs field during 
inflation as will be discussed in more detail in Sec. 4.3.

The temporal and spatial components of the above give,

\begin{eqnarray}
\partial_\tau (\mbox{\boldmath\bf $\nabla\cdot Z$})-
\nabla^2 Z_\tau+a^2M_Z^2Z_\tau & = & 0
\label{v=0}\\
 & & \nonumber\\
\Box\mbox{\boldmath\bf $Z$}-
\mbox{\boldmath\bf $\nabla$}
(\partial_\tau Z_\tau-\mbox{\boldmath\bf $\nabla\cdot Z$})+
a^2M_Z^2\mbox{\boldmath\bf $Z$} & = & 0
\label{v=i}
\end{eqnarray}
where \mbox{$\nabla^2\equiv\partial_i\partial_i$} with \mbox{$i=1,2,3$}
is the Laplacian, \mbox{$\Box\equiv\partial_\tau^2-\nabla^2$} is 
the D'Alembertian and {\boldmath\bf $\nabla$} is the divergence or the 
gradient. Taking the derivative of (\ref{CFRWZ}) we obtain the integrability 
condition,

\begin{equation}
\partial_\tau Z_\tau-\mbox{\boldmath\bf $\nabla\cdot Z$}=
-2\,[\partial_\tau\ln(aM_Z)]Z_\tau
\label{intcond}
\end{equation}
in view of which we can recast (\ref{v=i}) as,

\begin{equation}
\,[\Box+(aM_Z)^2]\mbox{\boldmath\bf $Z$}
+2\,[\partial_\tau\ln(aM_Z)]\mbox{\boldmath\bf $\nabla$}Z_\tau=0
\label{v=j}
\end{equation}

We now switch to momentum space by defining,

\begin{equation}
Z_\mu(\mbox{\boldmath $x$},\tau)= 
\int\frac{d^3k}{(2\pi)^3}\,{\cal Z}_\mu(\mbox{\boldmath\bf $k$},\tau) 
\exp(i\mbox{\boldmath\bf $k\cdot x$})
\end{equation}

Then (\ref{v=0}) and (\ref{v=j}) transform into,

\begin{eqnarray}
{\cal Z}_\tau+
\frac{i\partial_\tau(\mbox{\boldmath\bf $k\cdot{\cal Z}$})}{k^2+(aM_Z)^2}
 & = & 0
\label{calZ1}\\
 & & \nonumber\\
\,[\partial_\tau^2+k^2+(aM_Z)^2]\mbox{\boldmath $\cal Z$}+
2i\,[\partial_\tau\ln(aM_Z)]\mbox{\boldmath\bf $k$}Z_\tau & = & 0
\label{calZ2}
\end{eqnarray}
which, when combined, result in,

\begin{equation}
\,[\partial_\tau^2+k^2+(aM_Z)^2]\mbox{\boldmath $\cal Z$}+
\frac{2\,[\partial_\tau\ln(aM_Z)]}{k^2+(aM_Z)^2}\;
\partial_\tau
[\mbox{\boldmath\bf $k$(\mbox{\boldmath\bf $k\cdot{\cal Z}$})]}=0
\label{calZ}
\end{equation}
where \mbox{$k^2=$ {\boldmath $k\cdot k$}}.
We can decompose the gauge field into longitudinal and transverse modes
in the manner,

\begin{equation}
\mbox{\boldmath $\cal Z$}^{\parallel}\equiv
\frac{\mbox{\boldmath $k$}(\mbox{\boldmath $k\cdot{\cal Z}$})}{k^2}
 \hspace{2cm}
\mbox{\boldmath $\cal Z$}^{\perp}\equiv
\mbox{\boldmath $\cal Z$}-\mbox{\boldmath $\cal Z$}^{\parallel}
\label{longtrans}
\end{equation}

Then (\ref{calZ}) gives,

\begin{eqnarray}
\Big\{\partial_\tau^2+
\frac{2\,[\partial_\tau\ln(aM_Z)]k^2}{k^2+(aM_Z)^2}\;
\partial_\tau+k^2+(aM_Z)^2\Big\}
\mbox{\boldmath $\cal Z$}^{\parallel}
 & = & 0\\
 & & \nonumber\\
\Big[\partial_\tau^2+k^2+(aM_Z)^2\Big]
\mbox{\boldmath $\cal Z$}^{\perp}
 & = & 0
\label{Z}
\end{eqnarray}

In the following we will concentrate on the transverse component 
since the longitudinal component is really relevant only when interactions 
are taken into account, which, however, are not dealt with in this section.
For simplicity, we will drop the $^\perp$ symbol.

\subsection{Gravitational production of Z bosons}

In CFRW coordinates the scale factor during inflation and radiation 
domination is given by,

\begin{equation}
\begin{array}{cclr}
a(\tau) & = & -1/H\tau & \qquad -\infty<\tau\leq - 1/H\\
 & & & \\
a(\tau) & = & H\tau & \qquad  \tau\geq 1/H
\end{array}
\label{atau}
\end{equation}
such that \mbox{$a(-1/H)=a(1/H)=1$} and 
\mbox{$a^\prime(-1/H)=a^\prime(1/H)=H$}, where \mbox{$H\simeq$ const.} 
is the Hubble parameter during inflation.
Note that we have assumed a sudden transition from inflation to radiation.
If there is an intermediate era evolving like a matter-dominated Universe
this can be taken into account;
but we do not expect that the conclusions reached below will be affected by 
that, in that the spectrum should remain unaltered. Furthermore, because 
we are considering a GUT-scale inflationary model, we expect that 
reheating restores the EW symmetry and, therefore,  
\mbox{$M_Z(\tau\!>\!H^{-1})=0$}. In view of the above the equation for the
transverse component of \mbox{${\cal Z}_k(\tau)
\equiv|${\boldmath ${\cal Z}$}$^\perp(${\boldmath $k$},$\tau)|$} 
given by (\ref{Z}) becomes,\footnote{Since the coefficients in (\ref{Z}) for 
\mbox{{\boldmath ${\cal Z}$}$^\perp$} do not depend on direction,
it is reasonable to assume that (on average) the dependence is only on
the magnitude of the momentum, \mbox{$k=|${\boldmath $k$}$|$}.}

\begin{eqnarray}
\left(\partial_\tau^2+k^2+\frac{M^2_Z}{H^2}\frac{1}{\tau^2}\right) 
{\cal Z}_k(\tau) & = & 0 \qquad {\rm (inflation)}
\label{inf}\\
\left(\partial_\tau^2+k^2\right) 
{\cal Z}_k(\tau) & = & 0 \qquad {\rm (radiation)},
\label{rad}
\end{eqnarray}

As we will show in Sec. \ref{43},
$M_Z^2$ is created by the self-interaction of the EW Higgs-field. Because of 
this there is a slight logarithmic correction due to the time 
dependence\footnote{In fact, $1/\tau^2$ becomes $-\ln(-H\tau)/\tau^2$ because
\mbox{$M_Z^2\propto-\ln(-\tau H)$}.} of $M_Z$. 
However, we expect that this change does
not affect in any significant manner the results presented below.

Equations (\ref{inf}) and (\ref{rad}) have solutions in terms of Bessel
and harmonic functions, respectively. In particular (\ref{inf}) may be written
as,

\begin{equation}
\left(\partial_\tau^2+k^2-\frac{\nu^2-1/4}{\tau^2}\right){\cal Z}_k(\tau)=0
\label{bessel}
\end{equation}
with

\begin{equation}
\nu\equiv\sqrt{\frac{1}{4}-\frac{M^2_Z}{H^2}}
\label{nu}
\end{equation}
which has solutions in terms of the Hankel functions $H^{(j)}_\nu$, which 
have the Wronskian normalization 
\mbox{$W[H^{(1)}_\nu(z),H^{(2)}_\nu(z)]=
H^{(1)}_\nu(z) \partial_z H^{(2)}_\nu(z) - 
H^{(2)}_\nu(z) \partial_z H^{(1)}_\nu(z)=-4i/(\pi z)$}.
Hence, as a basis for the vector-field mode functions during inflation, we 
shall choose,

\begin{equation}
{\cal Z}_\nu^{(j)}(k\tau)=\frac{1}{2}\sqrt{-\pi\tau V}\,
H^{(j)}_\nu(-k\tau)\qquad (j=1,2) ,
\label{Zeigen}
\end{equation}
which solve (\ref{inf}) and have the Wronskian normalization \cite{as}
\begin{equation}
W[{\cal Z}^{(1)}_\nu(k\tau),{\cal Z}^{(2)}_\nu(k\tau)] 
\equiv {\cal Z}^{(1)}_\nu(k\tau)\partial_\tau{\cal Z}^{(2)}_\nu(k\tau)
-{\cal Z}^{(2)}_\nu(k\tau)\partial_\tau{\cal Z}^{(1)}_\nu(k\tau)
= Vi
\label{wronsk}
\end{equation}

Asymptotic forms of these solutions are easily found \cite{as}:

\begin{equation}
{\cal Z}_\nu^{(j)}(k\tau\!\rightarrow\!-\infty)\simeq
\sqrt{V/2k}\;\mbox{\large $e$}^{-i(-1)^j 
\left[k\tau+\frac{\pi}{2}(\nu+\frac{1}{2})\right]}
\left\{1+i(-1)^j\frac{4\nu^2-1}{8k\tau}+{\cal O}[(-k\tau)^{-2}]\right\}
\qquad (j=1,2)
\label{asyminfty}
\end{equation}
and also \cite{Arfken},

\begin{eqnarray}
{\cal Z}_\nu^{(j)}(k\tau\!\rightarrow\!0^{-}) & \simeq & 
\frac{1}{2}\sqrt{-\pi\tau V}\left[i(-1)^j \frac{\Gamma(\nu)}{\pi}
\left(-\frac{2}{k\tau}\right)^{\nu}\hspace{-0.2cm}+
\frac{1 - (-1)^ji\cot(\pi\nu)}{\Gamma(\nu+1)}
\left(-\frac{k\tau}{2}\right)^{\!\nu}\,\right]+\nonumber\\
 & & +\;{\cal O}[(-k\tau)^{2-\nu}]~,
\qquad \mbox{Re}[\nu]>0 \quad (j=1,2)
\label{asym0}
\end{eqnarray}

In the radiation era the eigenfunctions that solve (\ref{rad}) are

\begin{equation}
{\cal Z}_{\rm rad}^{(j)}(k\tau)=\sqrt{V/2k}\; e^{i(-1)^jk\tau}
\qquad (j=1,2) 
\label{Zrad}
\end{equation}
which are appropriately normalized as in (\ref{wronsk}).

It is now obvious that there is particle production since 
the eigenfunctions in inflation and radiation era are not orthogonal. 
A simple way of computing particle production is to do the matching 
at the end of inflation. Choosing for example ${\cal Z}^{(1)}_\nu$ 
in inflation, we can match it to the radiation era eigenfunctions as follows

\begin{eqnarray}
{\cal Z}_\nu^{(1)}(-k/H)
 & = & \alpha_k{\cal Z}_{\rm rad}^{(1)}(k/H)
+\beta_k{\cal Z}_{\rm rad}^{(2)}(k/H)\nonumber\\
 & & \label{match}\\
\partial_\tau{\cal Z}_\nu^{(1)}(-k/H)
 & = & \alpha_k\,\partial_\tau{\cal Z}_{\rm rad}^{(1)}(k/H)
+\beta_k\,\partial_\tau{\cal Z}_{\rm rad}^{(2)}(k/H)\nonumber
\end{eqnarray}
where $\alpha_k$ and $\beta_k$ are the Bogoliubov coefficients.
The asymptotic form of ${\cal Z}^{(1)}_\nu$ in (\ref{asyminfty}) implies that
the Bogoliubov transformation is diagonal in momentum space, and that the
coefficients depend only on the magnitude of the momentum 
$k=|\mbox{\boldmath $k$}|$. A similar transformation can be constructed for
${\cal Z}^{(2)}_\nu$. The fact that $\beta_k\neq 0$ implies that there is
particle production in inflation. The number of particles produced per mode
is then of the order of \mbox{$n_k\sim |\beta_k|^2$}.  
Using the asymptotic form (\ref{asym0}) for ${\cal Z}_\nu^{(1)}$ 
at the end of inflation, we can solve (\ref{match}) for $k\ll H$:

\begin{eqnarray}
\left(\begin{array}{c} \alpha_{{k}}\\*\beta_{{k}}
\end{array}\right)
& = &  e^{\pm ik/H}\left\{
\frac{\Gamma(\nu)}{4\sqrt{\pi}} \left[\mp
\left(\frac{1}{2}-\nu\right)\left({\frac{k}{2H}}\right)^{-\frac{1}{2}-\nu}
- 2 i \left({\frac{k}{2H}}\right)^{\frac{1}{2}-\nu}\right]\right.\;+
\label{bogo0}\\
& + & \!\!\left.\frac{\sqrt{\pi}(i- \cot\nu\pi)}{4\Gamma(1+\nu)}
 \left[\mp
\left(\frac{1}{2}+\nu\right)\left({\frac{k}{2H}}\right)^{\!\!-\frac{1}{2}+\nu}
\hspace{-0.3cm}-2i\left({\frac{k}{2H}}\right)^{\!\!\frac{1}{2}+\nu}\right]
\right\}+{\cal O}[(k/2H)^{\frac{3}{2}-\nu}]\nonumber\\
\end{eqnarray}

Note that the above Bogoliubov coefficients (\ref{bogo}) satisfy the
relation \mbox{$|\alpha_k|^2-|\beta_k|^2=1$}. 

We will consider the case when \mbox{$\nu=(H^2-4M_Z^2)^{1/2}/2H>0$}, 
i.e. $\nu$ is real and positive, which corresponds to \mbox{$M_Z<H/2$}.
In the limit \mbox{$M_Z\rightarrow H/2$} (i.e. \mbox{$\nu\rightarrow 0$}) 
conformal invariance is recovered and there is no particle 
production.\footnote{This is because, for \mbox{$M_Z\gsim H/2$} the 
quantum fluctuations of $Z$ are damped before reaching the horizon.}
Thus, from (\ref{bogo0}), to lowest order in $k/H$, we obtain,

\begin{equation}
|\alpha_k|^2\approx |\beta_k|^2\simeq\frac{|\Gamma(\nu)|^2}{16\pi}
\left(\frac{2H}{k}\right)^{2\nu+1}\left(\nu-\frac{1}{2}\right)^2
\label{bogo}
\end{equation}

In order to compute the spectrum of created particles, we need to evaluate,

\begin{eqnarray}
|{\cal Z}_{\rm rad}^{(1)}|^2-|{\cal Z}_{\rm vac}^{(1)}|^2
 & = & \frac{V}{2k}\left(|\alpha_k|^2-1\right)
\nonumber\\
|{\cal Z}_{\rm rad}^{(2)}|^2
 & = & \frac{V}{2k}|\beta_k|^2
\label{subvac}
\end{eqnarray}
where we have subtracted the vacuum contribution. 
In view of (\ref{bogo}) we see that the spectrum of 
$|{\cal Z}|^2$ on superhorizon scales has a slope by $k^{-2\nu-1}$
enhanced in comparison to the vacuum spectrum. The enhancement is stronger
for small masses \mbox{$M_Z\ll H$}, for which 
\mbox{$\nu\simeq \frac{1}{2}-(M_Z/H)^2$} and hence,

\begin{equation}
|{\cal Z}(k)|^2=2|\beta_k|^2\frac{V}{2k}
\simeq \pi^2V\Big(\frac{H}{2\pi}\Big)^2
\Big(\frac{M_Z}{H}\Big)^4
\frac{1}{k^3}
\Big(\frac{k}{2H}\Big)^{2(M_Z/H)^2}
\propto 
k^{-3+2(M_Z/H)^2}
\label{Zk}
\end{equation}
where the prefactor of $2|\beta_k|^2$ actually appeared as 
\mbox{$(|\alpha_k|^2+|\beta_k|^2)-1$}.

Therefore, \mbox{$Z_{\rm rms}(\ell)\propto\ell^{\nu-1/2}\approx 
\ell^{-(M_Z/H)^2}$} (see also appendix \ref{appb}).
Thus, for \mbox{$M_Z<H/2$}, the resulting rms spectrum for $Z$ is 
almost scale invariant. Due to 
the \mbox{$(\nu-\frac{1}{2})^2$} factor in (\ref{bogo}) we see that the 
enhancement of the spectrum would be canceled if the gauge field would have 
been {\rm exactly} massless, in which case \mbox{$\nu=\frac{1}{2}$}. This, for
instance, is the case of the massless photon. However, if 
\mbox{$0<M_Z\ll H$} then the fact that 
\mbox{$|\nu-\frac{1}{2}|\simeq(M_Z/H)^2\ll 1$} is compensated in the above by 
considering superhorizon modes, for which \mbox{$H\gg k$}.

It should be noted here that in a similar way one can generate any effectively
massless\footnote{i.e. with mass $<H$.}
gauge field during inflation and obtain a similar superhorizon spectrum.
A case of particular interest is the possibility of applying the above 
mechanism directly on the photon field by introducing a photon mass term
which vanishes at the end of inflation \cite{letter}. Another 
realization of this scenario is by considering a negative coupling Hybrid 
Inflationary model such that the expectation value of the scalar field, which
is coupled to the inflaton, is non-zero during inflation and becomes zero
at the end of it. If one considers this scalar field to be coupled to the 
photon then the above mechanism is operative \cite{next}. However, we feel 
that using the $Z$-field is rather more natural since, in this case, one does 
not require any additional scalars or couplings apart from the inflaton and 
the SM fields.

\subsection{\boldmath The origin and nature of $M_Z$ during 
inflation}\label{43}

In computing the spectrum we have assumed that conformal invariance is 
broken by a mass term which is due to a condensate of the EW Higgs field.
Here we look in more detail at the underlying mechanism behind the formation
of such a condensate. To this purpose we need to recast (\ref{yfe}) 
in the Hartree approximation, which we describe in Appendix \ref{appa}. 
We have,

\begin{equation}
\partial_\tau^2y_k
+\Big[3\lambda_*\langle y^2\rangle_{_V}-a^2 M_{EW}^2
-g^2a^2 \sum_{j=+,-}\langle(W^j)^2\rangle_{_V} 
-g_Z^2a^2 \langle Z^2\rangle_{_V} 
-\frac{\partial_\tau^2 a}{a}\Big]y_k=0
\label{yk}
\end{equation}
where $g^2a^2 \langle\cdot\rangle_{_V}$ denotes the Hartree 
back reaction. Assuming that
\mbox{$\langle\Psi^\dagger\Psi\rangle_{_V}\equiv 
\langle y^2\rangle_{_V}/a^2$},
\mbox{$\langle(W^j)^2\rangle_{_V}$} 
and \mbox{$\langle Z^2\rangle_{_V}$} 
are slowly varying functions in time, 
during inflation (\ref{yk}) can be recast in the form of (\ref{bessel}) with,

\begin{equation}
\nu^2_*\equiv\frac{9}{4}+\frac{1}{H^2}\Big[
-3\lambda_*\langle\Psi^\dag\Psi\rangle_{_V}+M_{EW}^2
+g^2\sum_{j=+,-}\langle(W^j)^2\rangle_{_V} 
+g_Z^2\langle Z^2\rangle_{_V} 
\Big]
\label{nu*}
\end{equation}
where we used $\partial_\tau^2 a/a=2/\tau^2$. As is shown in \ref{a21}, 
the Hartree backreaction for the gauge fields gives
\mbox{$\langle Z^2\rangle_{_V}\simeq
\langle(W^j)^2\rangle_{_V}\lsim H^2$}. Thus, the
backreaction of the gauge fields is less than \mbox{$(2g^2+g_Z^2)H^2<H^2$}. 
Also \mbox{$M_{EW}^2\ll H^2$}. Therefore, the dominant contribution in
(\ref{nu*}) must come from the $3\lambda_*\langle\Psi^\dag\Psi\rangle_{_V}$
term. Now assuming that 
$\nu_*$ changes slowly (adiabatically) the solution of (\ref{yk})
is given by (\ref{Zeigen}),

\begin{equation}
y_{\nu_*}^{(j)}(k\tau)=\frac{1}{2}\sqrt{-\pi\tau V}\,H^{(j)}_{\nu_*}(-k\tau)
\qquad (j=1,2) ,
\label{yhank}
\end{equation}
where we used again the following normalization 
[{\em cf.} (\ref{wronsk})]
 
\begin{equation}
W[y^{(1)}_{\nu_*}(k\tau),y^{(2)}_{\nu_*}(k\tau)] =Vi
\label{wronsk3}
\end{equation}

Assuming that \mbox{Re$[\nu_*]>1/2$} then, in analogy with (\ref{asym0}), 
(\ref{yhank}) implies,

\begin{equation}
|y_{\nu_*}^{(j)}(k\tau)|^2
\stackrel{k\rightarrow 0}{\longrightarrow}
\frac{\Gamma(\nu_*)^2}{\pi}\frac{V}{2k}
\left(-\frac{2}{k\tau}\right)^{2\nu_*-1} ,
\label{yasym0}
\end{equation}
which, when subtracting the vacuum in the Hartree approximation
and considering \mbox{$\nu_*\simeq 3/2$}, results in an almost scale 
invariant spectrum because (See \ref{a22}),

\begin{equation}
\langle\Psi^\dag\Psi\rangle\simeq\frac{H^2}{4\pi^2} 
\ln|\frac{\tau_i}{\tau}|
\label{ylnt}
\end{equation}
where $\tau_i$ denotes the beginning of inflation\footnote{Remember that, 
during inflation \mbox{$a\propto 1/|\tau|$}}. 
The relevant quantity for the backreaction is [{\em cf.} (\ref{nu*})]
\mbox{$3\lambda_*\langle\Psi^\dag\Psi\rangle/H^2\propto -\ln|\tau|$},
which is a very weak function of time and thus, the adiabaticity assumption
is justified. Moreover, for \mbox{$\lambda_*<1$}, we see from (\ref{nu*})
that indeed \mbox{$\nu_*\simeq 3/2$} as assumed. 

In view of (\ref{MZ}), the above considerations suggest that,\footnote{
Considering (\ref{atau}) we see that,
\mbox{$M_Z\simeq g_Z(H/2\pi)\sqrt{\Delta N}$}
where \mbox{$\Delta N=N_{\rm tot}-N$} and we used \mbox{$a=e^N$}, with $N$ 
being the number of e-foldings remaining until the end of inflation and 
$N_{\rm tot}$ being the total number of e-foldings. Thus (\ref{temp}) may be
understood roughly as follows: Every e-folding the gravitationally generated
fluctuation over the horizon volume of the EW-Higgs field is of the order of 
the Gibbons-Hawking temperature \mbox{$\delta\psi_k\simeq H/2\pi$}. 
The quantity $\langle\Psi^\dag\Psi\rangle$ represents
an accumulated ``memory'' of these fluctuations which ``pile up'' while
inflation continues. Thus, due to the stochastic nature of the fluctuations 
they perform a random walk, which, after $\Delta N$ steps is responsible for
the $\sqrt{\Delta N}$ increase.}

\begin{equation}
M_Z^2\simeq g_Z^2\langle\Psi^\dag\Psi
\rangle\simeq\frac{g_Z^2}{4\pi^2}H^2 
\ln|\frac{\tau_i}{\tau}|
\label{temp}
\end{equation}

\subsection{The generated hypermagnetic field}

Because reheating introduces large temperature corrections to the scalar 
potential (\ref{u}) the EW-symmetry gets restored and the hypercharge 
field becomes massless. From (\ref{ZWY}) and (\ref{AWY}) we obtain,

\begin{equation}
Y_\mu=\cos\theta_WA_\mu-\sin\theta_WZ_\mu
\label{YAZ}
\end{equation}

However, in contrast to the $Z$-boson case, the superhorizon spectrum of the 
photon is not almost scale invariant but, in fact, scales as 
\mbox{$A_{\rm rms}(\ell)\propto \ell^{-1}$},
because conformal invariance is retained for the photon during inflation.
Thus, the contribution of the photon to the superhorizon spectrum of the
hypercharge field is negligible compared to the one of the $Z$-boson. 
Therefore, the hypercharge superhorizon spectrum at the end of inflation
is also almost scale invariant,

\begin{equation}
|Y_\mu|\simeq\sin\theta_W |Z_\mu|
\label{Yrms}
\end{equation}

The hypercharge field during the radiation era is massless and Abelian and,
therefore, it satisfies the equivalent of Maxwell's equations. 
The associated hypermagnetic field is defined as
\mbox{{\boldmath $B$}$^Y\equiv\mbox{\boldmath $\nabla$}\times 
\mbox{\boldmath $Y$}$}. As estimated in the Appendix~\ref{appb},
the rms value over a given superhorizon scale $\ell$ 
of the hypermagnetic field due to the superhorizon spectrum of the 
$Z$-boson field is,

\begin{equation}
[B_{\rm rms}^Y(\ell)]^2=\frac{18(\sin\theta_W)^2}{\pi^2\ell^6}
\int_0^{+\infty} dk\Big[\frac{1}{k}\sin(k\ell)-\ell\cos(k\ell)\Big]^2
\frac{|{\cal Z}(k)|^2}{V}
\label{Brms0}
\end{equation}

Inserting (\ref{Zk}) into the above and after some algebra 
(see appendix \ref{appb}) we obtain, 

\begin{equation}
B_{\rm rms}^Y(t_{\rm end})=\frac{3\sqrt{2}\sin\theta_W}{8\pi^2}
\Big(\frac{M_Z}{H_\ell}\Big)^2k_{\rm end}H_{\rm end}
\label{Brms}
\end{equation}
where $H_\ell$ is the Hubble parameter at the time the relevant scale exits 
the Horizon during inflation, 
\mbox{$k_{\rm end}=k(t_{\rm end})=2\pi/\ell$} is the physical momentum at the 
end of inflation, because in (\ref{atau}) the scale factor $a$ has been 
normalized at that time and\footnote{For 
pure de-Sitter inflation we have \mbox{$H_\ell=H_{\rm end}$}.}
\mbox{$H_{\rm end}=H(t_{\rm end})$}.

\subsection{Evolution of the magnetic field}

The creation of a thermal bath of SM particles at reheating 
freezes the hypermagnetic field into
the reheated plasma in a way analogous to regular magnetic fields. 
At this point we should mention that, apart from the hypercharge component
the $W$-boson fields also have an almost scale invariant spectrum over 
superhorizon scales at the end of inflation, generated gravitationally 
because their conformal invariance is broken during inflation in the same 
manner as in the case of the $Z$-boson. After reheating, however, because 
the $W$ gauge fields are non-Abelian they are screened and decay due to the
development of a magnetic mass \cite{mb}, \mbox{$M_B\approx$ 0.28 $g^2T$}. 
Because of this we will focus on the evolution of the hypercharge magnetic 
field configuration.

\input epsf

\begin{figure}
\begin{center}
\leavevmode
\hbox{\epsfxsize=4.9in
\epsffile{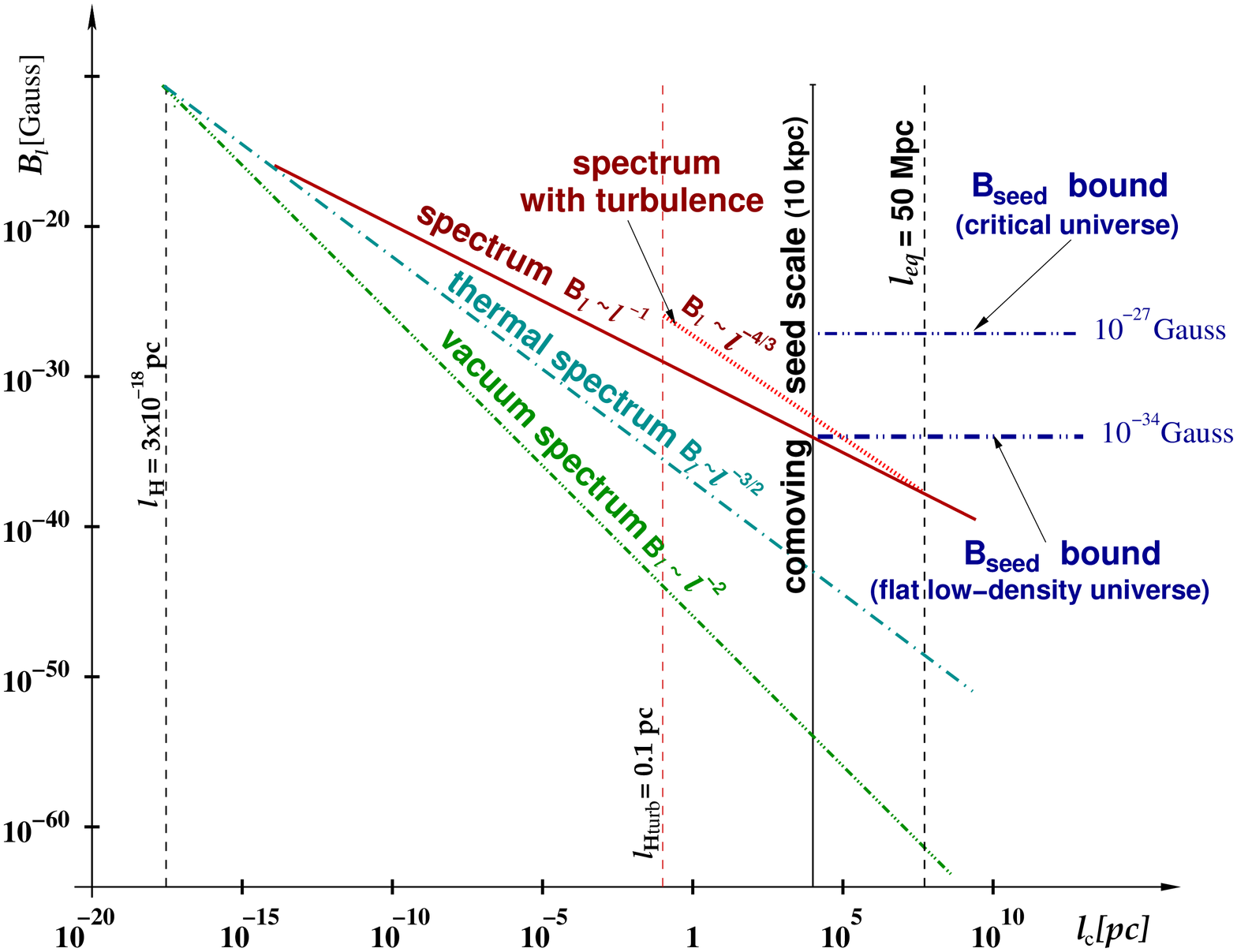}}
\end{center}
\caption{\footnotesize
Comoving magnetic field spectra and relevant seed field bounds, plotted 
against the comoving scale~$\ell _c$.
The steepest line (dash-dot-dot-dot) depicts the (unsubtracted) 
vacuum spectrum \mbox{$B_{\rm vac}(\ell)\propto\ell^{-2}$}.
At the comoving scale \mbox{$\ell_c=10$ kpc} one has 
\mbox{$B_{\rm vac}^{\rm com}(10$ kpc$)\sim 10^{-54}$Gauss}. 
The line of intermediate 
steepness (dash-dot) corresponds to the thermal spectrum
\mbox{$B_{\rm th}(\ell)\propto\ell^{-3/2}$}, which meets the vacuum spectrum 
at the comoving scale $\ell_H$ corresponding to the Horizon at the end of 
inflation. The solid line is the primordial magnetic field, resulting from 
inflation through our mechanism, 
\mbox{$B_{\rm rms}(\ell)\propto\ell^{\nu-3/2}\approx\ell^{-1}$}, 
as calculated by (\ref{Bo}). On the scale \mbox{$\ell_c=10$ kpc} we find 
\mbox{$B_{\rm rms}^{\rm com}(10$ kpc$)\sim 10^{-34}$Gauss}, 
i.e. about twenty orders of magnitude stronger than the value 
corresponding to the vacuum spectrum. 
This is to be compared with the dynamo bounds rescaled 
by a factor $10^{-4}$, 
corresponding to the collapse-amplification enhancement factor
and the factor due to the scaling between galaxy formation and 
the present (see main text). These bounds read
\mbox{$B_{\rm seed}\geq 10^{-27}$Gauss} 
for a Universe with critical matter density and 
\mbox{$B_{\rm seed}\geq 10^{-34}$Gauss}
for a Universe dominated
by dark energy (flat, low-density Universe). We also show (dotted line) the
spectrum enhanced by helical turbulence (at \mbox{$\ell_c=10$ kpc} an 
enhancement of about 20 is obtained). 
Finally, we note that we expect our primordial magnetic field 
spectrum to depart from the $\ell^{-1}$ scaling law near the scale $\ell_H$
and approach approximately the thermal spectrum since 
\mbox{$\nu\rightarrow 0$} towards the end of inflation. In that way the 
spectrum of our magnetic field joins the vacuum spectrum at $\ell_H$.}
\label{results}
\end{figure}

The physical momentum scales as \mbox{$k\propto a^{-1}$}. 
Thus, by scaling \mbox{$k_{\rm end}$} to the present day we find,

\begin{equation}
k_{\rm end}=\frac{2\pi}{\ell_c}(\frac{T_{\rm reh}}{T_{\sc cmb}})
\label{k}
\end{equation}
where $\ell_c$ is the corresponding scale of the mode in question today, 
$T_{\rm reh}$ is the temperature resulting from prompt reheating, 
$T_{\sc cmb}$ is the temperature of the CMBR at present and we have used 
that \mbox{$a\propto T^{-1}$} at all times. 

If we assume that the field remained frozen until today then, because
flux conservation requires \mbox{$B\propto a^{-2}$}, we find that
the magnitude of the magnetic field today would have been,

\begin{equation}
B^{\rm com}_{\rm rms}=\cos\theta_W\,B^Y_{\rm rms}(t_{\rm end})\,
(\frac{T_{\sc cmb}}{T_{\rm reh}})^2
\label{B}
\end{equation}
where $\cos\theta_W$ is introduced due to the projection of the 
hypercharge onto the photon at the electroweak transition according to 
(\ref{AWY}).

Using (\ref{Brms}) and (\ref{k}) the above gives, 

\begin{equation}
B^{\rm com}_{\rm rms}=\frac{1}{4}(\frac{3g_*}{40\pi^2})^{1/4}
\sin(2\theta_W)
(\frac{T_{\sc cmb}}{\ell_c})
(\frac{V_{\rm end}^{1/4}}{m_P})
(\frac{M_Z}{H_\ell})^2
\label{B0}
\end{equation}
where, for prompt reheating, \mbox{$V_{\rm end}\simeq
(\pi^2/30)g_*T_{\rm reh}^4$}, with \mbox{$g_*=106.75$}
being the number of relativistic degrees of freedom at the end of 
inflation.\footnote{Note that $g_*$ can be somewhat larger in simple 
extensions of the SM. For example in the Minimal Supersymmetric SM, 
\mbox{$g_*=229$}. Thus, \mbox{$g_*=106.75$} may be viewed as a lower 
bound to the actual value of $g_*$.} 
Putting the numbers in (\ref{B0}) one finds,

\begin{equation}
B^{\rm com}_{\rm rms}=5\times 10^{-33}
\left(\frac{\mbox{1 Mpc}}{\ell_c}\right)
\left(\frac{M_Z}{H_\ell}\right)^2
\frac{V_{\rm end}^{1/4}}{m_P}
\;\mbox{Gauss}
\label{Bo}
\end{equation}
where \mbox{$T_{\sc cmb}=2.37\times 10^{-13}$ GeV} and
\mbox{$\sin(2\theta_W)\approx 0.84$}.
The above corresponds to {\em the spectrum of the 
primordial magnetic field as it would have been today were there no 
galactic collapse and subsequent dynamo amplification}. Typically, such a 
field is referred to as ``comoving''. The comoving spectrum of our primordial 
field, as given by (\ref{Bo}) is shown in Fig.~\ref{results}, where the 
substantial amplification compared with the vacuum spectrum is apparent.

However, the actual physical field, being frozen into the plasma, will be 
affected by the gravitational collapse during structure formation. Since we 
are interested in seeding the dynamo, we may use (\ref{B0}) to estimate the 
seed field at the time of galaxy formation. Scaling back the comoving field to
galaxy formation provides an amplification factor of 
\mbox{$(1+z_{\rm gf})^2$}, where $z_{\rm gf}$ is the redshift that corresponds 
to galaxy formation. This is due to the expansion of the Universe 
(viewed backwards). Moreover, the collapse of matter into galaxies 
brings about a further amplification 
of magnetic fields by a factor given by the fraction of the galactic matter 
density to the matter density of the Universe at galaxy formation, 
\mbox{$(\rho_{\rm gal}/\rho_{\rm gf})^{2/3}$}, where 
\mbox{$\rho_{\rm gf}=\rho_0 (1+z_{\rm gf})^3$} with $\rho_0$ being the 
matter density of the Universe at present. The above result in an overall 
amplification factor of about $\sim 10^4$.
Taking this into account and considering 
that the scale of the largest turbulent eddy corresponds to the comoving
scale of about \mbox{$\ell_c\simeq 10$ kpc} before the gravitational collapse 
of the protogalaxy, we find, 

\begin{equation}
B_{\rm seed}\sim 10^{-30}\mbox{Gauss}
\label{B1}
\end{equation}
where we assumed GUT-scale inflation \mbox{$V_{\rm end}\sim 10^{16}$GeV} and 
we have used that, for total number of inflationary e-foldings $\sim$~100 we 
have, \mbox{$(\frac{M_Z}{H_\ell})^2\sim 0.01$}.
The above seed field is sufficient to trigger the galactic dynamo in the case 
of a spatially-flat, dark-energy (e.g. a cosmological constant or quintessence) 
dominated Universe \cite{dlt}.

A supplementary increase in field strength is obtained if we assume that
the magnetic field does not freeze into the plasma upon creation, but
rather that its correlation length grows quicker than the scale factor,
as is the case for helical turbulence \cite{Son}. Such a causal mechanism can
only operate on a given comoving scale after this scale has reentered the
horizon. One can show that the growth of correlations
due to turbulent evolution leads to an additional amplification of about
\mbox{$(\ell_{\rm eq}/\ell_c)^{2(1-\nu)/3}$},
where \mbox{$\ell_{\rm eq}\sim 50$ Mpc} denotes the equal matter-radiation
horizon today. For \mbox{$\ell_c=10$ kpc} and \mbox{$\nu\simeq 1/2$}
the amplification factor is about 20.

In the above we have assumed prompt reheating. However, it is possible that 
reheating may be rather inefficient and, therefore, a long-lasting process,
especially in the case when the inflaton decays prominently into bosons 
\cite{charts2}. Indeed, the reheating temperature is typically given by, 
\mbox{$T_{\rm reh}\lsim\sqrt{\Gamma\,m_P}$}, where $\Gamma$ is the 
perturbative decay rate of the inflaton \cite{charts2}\cite{charts3}. 
If the inflaton decays into bosons then $\Gamma$ 
decreases with time and it is possible that the decay of the inflaton 
particles (inflatons) ceases before being completed leaving a small fraction 
of the inflatons as dark matter \cite{charts2}\cite{charts3}. 
In such case $T_{\rm reh}$ may be rather low 
and can satisfy the gravitino constraint (\mbox{$T_{\rm reh}\lsim 10^9$GeV}) 
even for GUT-scale inflation. Of course, an initial stage of preheating may 
increase substantially the overall efficiency of reheating \cite{stocha}. 
However, in SUSY-HI, since \mbox{$\lambda\sim h$}, preheating is chaotic and 
likely to be also rather inefficient \cite{bellido}. Finally, one can think of 
variations similar to ``smooth Hybrid Inflation'' of \cite{rachel2}, which may 
achieve a rather low reheating temperature. 
Still, as shown in (\ref{Brms}), the amplitude of our hypermagnetic field is, 
in fact, decided by the inflationary energy scale and not by the reheating 
temperature. Indeed, it is easy to show that, even in the case of 
long-lasting, inefficient, reheating the strength of our seed field at galaxy 
formation remains essentially unmodified, so that (\ref{Bo}) is still 
applicable. Therefore, our mechanism may manage to 
satisfy the gravitino overproduction constraint while still able to provide a 
sufficiently strong seed field. Thus, our SUSY-HI inflationary model can be 
incorporated in a Supergravity framework without problem.

\input epsf

\begin{figure}
\begin{center}
\leavevmode
\hbox{\epsfxsize=3.0in
\epsffile{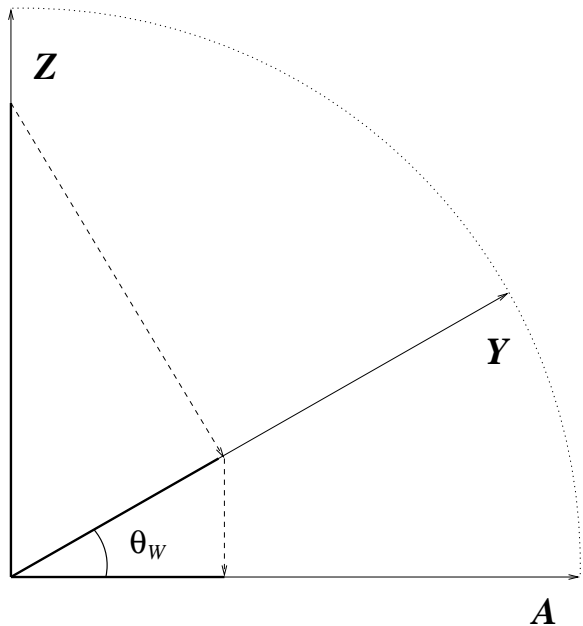}}
\end{center}

\caption{\footnotesize
The formation of a magnetic field from the generation of the $Z$-boson field
has to pass through the intermediate hypercharge stage similarly to 
light when crossing a set of orthogonal polarizers, which needs an 
intermediate third polarizer at some angle $\theta$ in order to go through.}
\label{polarisers}
\end{figure}

At first sight, it may seem unlikely that a magnetic field is obtained
from the gravitational production of the $Z$-boson field, which is orthogonal 
to the photon field. In fact, the situation is analogous to that of light 
polarizers (see Fig.~\ref{polarisers}). With two
orthogonal light polarizers no light passes. When a third polarizer
is inserted at an angle $\theta$ (with respect to the second polarizer)
however, some of the photons do pass. The photon amplitude is reduced
by $\frac{1}{2}\sin(2\theta)$, just as in the case of the $Z$-field.
The main advantage of the amplification mechanism presented here
is its naturalness. Indeed, no fields are required except those of the 
standard model and the inflaton. Moreover, the mechanism is
independent of the particular model of inflation considered, but can be 
thought of as a generic consequence of inflationary theory itself.\footnote{ 
Note that most of the successful inflationary models (e.g. chaotic, natural, 
hybrid) have GUT-scale $V_{\rm end}$.}

An exciting possibility is that the above magnetic field can be amplified 
even further during preheating in an explosive resonant way. In what follows 
we explore this possibility using the particular model of Flipped~SU(5)
SUSY-HI.

\section{Preheating}

In addition to the growth of the superhorizon spectrum due to the 
breaking of the $Z$-conformal invariance in the inflationary period, the 
generated hypermagnetic field may be amplified more during preheating 
due to the backreaction of the hypercharge source current. From
(\ref{JYH}) it is evident that the growth of the amplitudes of the 
GUT-massive gauge fields $X^\alpha_\mu$ and $Y^\alpha_\mu$ is a prerequisite
for this source current to be substantial. In Hybrid Inflation this is 
achieved through their mass term given by (\ref{masses}), which resonantly 
pumps energy from the GUT-Higgs field. The latter is resonantly produced at 
the end of inflation due to its coupling with the oscillating inflaton.

\subsection{Parametric resonance in Hybrid Inflation}

\subsubsection{The mode equations of the scalar fields}

We shall consider production of the supermassive vector bosons 
by parametric resonance caused by oscillations of $s$ and $\phi$. 
Since $s$ is a singlet and does not couple to the vector fields, 
the production of gauge fields is due to the GUT-Higgs field. 
A significant production of modes of any kind occurs only after
$\phi$ and $s$ start oscillating around the bottom of the potential. 
Oscillations of $\phi$ are not harmonic, but to a good 
approximation they are periodic, with an almost constant period. 
This is so since the Hubble constant is small,
\mbox{$H=\sqrt{(\rho_{\rm inf}/3)}/m_P\sim\lambda M^2/m_P\sim 10^{13}$GeV},
while a typical oscillation frequency is \mbox{$\omega\sim\sqrt{\lambda}\,M$},
where we have used (\ref{V}) to estimate 
\mbox{$\rho_{\rm inf}\equiv V(0,s_c)$}.

Since, typically, the resonant decay of a field lasts a few hundreds of 
oscillations, the decay time of the inflaton is \mbox{$\lsim H^{-1}$}. 
On these grounds one can neglect the expansion of the Universe during 
preheating and regard the scale factor to be approximately constant
\mbox{$a\simeq 1$} in agreement with the normalization (\ref{atau}).
Therefore, the field equations (\ref{cfrw-s}) and (\ref{cfrw-f}) become,

\begin{eqnarray}
\Big(\partial_\tau^2+
h\varphi^2\Big)\sigma+\frac{\partial V_\sigma}{\partial\sigma} & = & 0\\
 & & \nonumber\\
\Big[\partial_\tau^2+
\lambda(\varphi^2-M^2)+h\sigma^2\Big]\varphi=0
\label{above}
\end{eqnarray}
where in (\ref{above})
we have ignored the backreaction of the supermassive GUT-bosons because
at the onset of preheating their amplitude is negligible.
In terms of the field \mbox{$\bar{\varphi}=\varphi-M$} the above can be 
rewritten as,

\begin{eqnarray}
\Big[\partial_\tau^2+
\lambda\left(\bar{\varphi}^2+3\bar{\varphi}M +2M^2\right)
  +h\sigma^2\Big]\bar{\varphi} + h\sigma^2M
   &=& 0
\nonumber\\
\Big[
\partial_\tau^2+
h(\bar{\varphi}^2+2\bar{\varphi}M +M^2) \Big]\sigma
 &=& 0 .
\label{scal}
\end{eqnarray}
where we have ignored the term $\frac{\partial V_\sigma}{\partial\sigma}$ 
since the slope of the $V_\sigma$ potential is negligible compared to the 
interaction term. The solutions to these equations are damped periodic 
oscillatory functions\footnote{We do not consider here the effects of chaotic 
resonant behavior~\cite{bellido}, which 
may occur in hybrid inflationary models.}
which in general have the following form,

\begin{eqnarray}
\bar{\varphi}  &=& \varphi_0 P_\varphi(\omega_\varphi\tau)
\nonumber\\
\sigma &=& \sigma_0 P_\sigma(\omega_\sigma\tau) ,
\label{scalamp}
\end{eqnarray}
where  \mbox{$P_\varphi(x+2\pi)=P_\varphi(x)$}, and 
\mbox{$P_\sigma(x+2\pi)=P_\sigma(x)$},  
\mbox{max$\{|P_\varphi|\}=1$} and \mbox{max$\{|P_\sigma|\}=1$}, and 
$\varphi_0$ and $\sigma_0$ are the amplitudes of oscillations, which are in 
principle time-dependent since $\varphi$ and $\sigma$ eventually decay 
and redshift.

The corresponding mode equations can be obtained with Fourier transform.
The result is, 

\begin{eqnarray}
\Big[\partial_\tau^2
+ k^2+
\lambda\left(3\bar{\varphi}^2+6\bar{\varphi}M +2M^2\right)
  +h\sigma^2\Big]\varphi_k 
+2h\sigma(\bar{\varphi} + M)\sigma_k
   &=& 0\qquad
\label{phimode}
\\
 & & \nonumber\\
\Big[
\partial_\tau^2 + k^2+
 h(\bar{\varphi}^2+2\bar{\varphi}M +M^2) 
\Big]\sigma_k
 + 2h\sigma(\bar{\varphi} + M)\varphi_k &=& 0
\label{smode}
\end{eqnarray}

By making use of 
\mbox{$\langle\bar{\varphi}^2\rangle=\varphi_0^2/2$} and 
\mbox{$\langle\sigma^2\rangle=\sigma_0^2/2$}, we can easily estimate the 
effective masses of the modes to be,

\begin{eqnarray}
m_\varphi^2 &=& 2\lambda\left(M^2 + \frac{3}{4}\varphi_0^2\right)
  +\frac{1}{2}h\sigma_0^2 \simeq 4\lambda M^2\nonumber\\
 & & \label{effmasses}\\
m_\sigma^2  &=&
 h\left(M^2+\frac{1}{2}\varphi_0^2\right) 
 \simeq \frac{3}{2}hM^2 
\nonumber
\end{eqnarray}
so that the dispersion relations read,

\begin{eqnarray}
\omega_\varphi^2 &=&
k^2+2\lambda\left(M^2+\frac{3}{4}\varphi_0^2\right)+\frac{1}{2}h\sigma_0^2
  \simeq k^2 +  4 \lambda M^2 \nonumber\\
 & & \label{disp}\\
\omega_\sigma^2  &=& k^2 +
 h\left(M^2+\frac{1}{2}\varphi_0^2\right)
 \simeq k^2 +  \frac{3}{2}h M^2 
\nonumber
\end{eqnarray}
where \mbox{$\varphi_0^2\simeq M$} and 
\mbox{$\sigma_0^2\simeq s_c^2=(\lambda/h)M^2$}.
Note that the effective mass of the $\bar{\varphi}$ modes is a factor of 
$\sqrt{2}$ larger than the Higgs mass \mbox{$M_H=\sqrt{2\lambda}\,M$}.

We need to investigate whether the mode equations (\ref{phimode}) and 
(\ref{smode}) result in a broad or narrow resonance, and also whether the 
leading (broad) resonance is active. To do that we have to consider the 
quality factors $q$ for the mode equations and then compare
them to the effective masses (\ref{effmasses}). 

\subsubsection{The characteristics of the resonance}

The generic form for a resonance equation is 

\begin{equation}
\frac{d^2\psi_k}{d^2\varpi}
+\Big[A_k+\sum_i2q_iP_i(\varpi)\Big]\psi_k =0
\hspace{1cm}
\mbox{where}
\hspace{1cm}
A_k =\frac{m_{\psi\,\rm eff}^2 + k^2}{\omega_\varphi^2} = 1 
\label{resequ}
\end{equation}
and $q_i$ are the quality factors with oscillating functions 
$P_i(\varpi)$ 
(\mbox{$\langle P_i\rangle_\varpi=0$}, 
\mbox{max$_x\{P_i\}=1$}) whose period is either $\pi$ or $2\pi$, and 
\mbox{$\varpi\equiv\omega_\varphi\tau$} is the rescaled 
time variable. In the Mathieu case, 
\mbox{$P_i=\cos(2\varpi)$}.
Some of the references where one can find instability charts, which 
are plots of the Floquet exponent $\mu$ in terms of $q$ and $A$, 
are listed in \cite{charts1}\cite{charts2}\cite{charts3}.

Consider first the quality factors for (\ref{phimode}),

\begin{equation}
q_{\varphi\, 1} = \frac{3\lambda\varphi_0^2}{4\omega_\varphi^2}
\sim \frac{3}{16}
\qquad
q_{\varphi\, 2} = \frac{6\lambda M\varphi_0}{2\omega_\varphi^2}
\sim \frac{3}{4}
\qquad
q'_{\sigma\, 1} = \frac{2hM\sigma_0}{2\omega_\varphi^2} 
\sim \frac{1}{4}\sqrt{\frac{h}{\lambda}} 
\label{qphi}
\end{equation}

Since \mbox{$\omega_\varphi^2\simeq 4\lambda M^2$}, we conclude that all the
quality factors of (\ref{phimode}) are generically of the order unity. 
The quality factors for (\ref{smode}) are on the other hand

\begin{equation}
q_{\sigma\, 1} = \frac{h\varphi_0^2}{4\omega_\varphi^2}
\sim \frac{1}{16}\frac{h}{\lambda} 
\qquad
q_{\sigma\, 2} = \frac{2hM\varphi_0}{2\omega_\varphi^2}
\sim  \frac{1}{4}\frac{h}{\lambda} 
\qquad
q'_{\varphi\, 1} = \frac{2hM\sigma_0}{2\omega_\varphi^2} 
\sim  \frac{1}{4}\sqrt{\frac{h}{\lambda}}
\label{qs}
\end{equation}
so that they are typically smaller than the quality factors (\ref{qphi}) of 
(\ref{phimode}).\footnote{Remember that in SUSY-HI \mbox{$\lambda=2h$}.} 
Further we can compute 

\begin{eqnarray}
A_\varphi(k\!=\!0) &=&
  \frac{m_\varphi^2}{\omega_\varphi^2}=1 \sim \frac{4}{3}q_{\varphi\, 2}
\nonumber\\
A_\sigma(k\!=\!0) &=& \frac{m_{\sigma,\bar{\varphi}=0}^2}{\omega_\varphi^2}
         = \frac{1}{4}\frac{h}{\lambda} \sim q_{\sigma\, 2}
\label{Aphis}
\end{eqnarray}
which then determines onto which resonance the field can decay. 
Note that since the position of the first resonance is at \mbox{$A\sim 1$} 
for \mbox{$q<1$}, we conclude that the $\varphi$ field tends to 
decay into the first resonance such that the infrared superhorizon modes get
populated (see figure in \cite{charts2}) with \mbox{$\mu\sim 0.3$} or smaller.
A similar statement applies to the $\sigma$ field.

The strength of the broad resonances may be as large as \mbox{$\mu\sim 0.2$} 
along the $A=2q$ line, and becomes rapidly stronger at smaller values of $A$,
i.e. \mbox{$A<2q$} (cf. Eq. (7) in \cite{charts1}). Even if the field can 
decay only into the second available resonance, the strength is expected to be 
of the order \mbox{$\mu_2\sim 0.04$}, 
just like it is in the case of the chaotic inflation with a quartic scalar
interaction term, which decays into the second resonance. 

Regarding the $\sigma$ field, depending on the choice of the couplings, 
$q_{\sigma\, 2}$ 
can be larger or smaller then unity. Generically, however,
we expect that \mbox{$\lambda\sim h$}, and hence  $q_{\sigma\, 2}$
is typically of the order (but smaller then) unity. 
In SUSY-HI we have \mbox{$\lambda=2h$} and, therefore, 
\mbox{$q_{\sigma\, 2}\simeq\frac{1}{8}<1$}. 

In the above, by taking \mbox{$a\simeq 1$} we have ignored Hubble damping. 
It is known that the latter, in fact, results in stochastic resonant 
amplification, in the sense that the action of the damping term may be
positive or negative on occasion, depending on whether the amplitude of the 
oscillating fields is increasing or decreasing. However, as has been shown in 
\cite{charts3}\cite{stocha}, the net effect of such a term is, in fact, in 
favor of resonant production.

This analysis has not covered the mixing terms in 
(\ref{phimode}) and (\ref{smode}). Numerical simulations
done by the authors of \cite{charts1} indicate that these terms typically 
cannot prevent resonant decay, and in some cases can enhance it. 

\subsection{Resonant production of massive vector bosons}

The field equations for the supermassive GUT-bosons are (\ref{feX}) and
(\ref{feYa}), while (\ref{feV}) is the one of for the $V$-boson. The resonant
production of the massive bosons is due to the fact that their equations of
motion are of harmonic form with frequencies given by their oscillating mass 
term.

\subsubsection{The parametric resonance equations}

At the onset of preheating we may ignore the backreaction source current 
because the initial amplitude of the massive bosons is negligible. Later
on the source current contributes additionally to the production of the 
massive bosons.

Let us concentrate on (\ref{feX}) first. Proceeding in a similar manner as in 
\ref{modequsZ} we 
obtain for the longitudinal and transverse modes,

\begin{eqnarray}
\Big[(\partial_\tau^2+k^2+M_{XY}^2)+
\frac{2(\partial_\tau\ln M_{XY})k^2}{k^2+M_{XY}^2}\partial_\tau\Big]
\mbox{\boldmath $\cal X$}^{\alpha\parallel}
 & = & 0\label{Xparal}\\
 & & \nonumber\\
\Big(\partial_\tau^2+k^2+M_{XY}^2
\Big)\mbox{\boldmath $\cal X$}^{\alpha\perp}
 & = & 0
\label{Xperp}
\end{eqnarray}
where 

\begin{equation}
X_\mu^\alpha(\mbox{\boldmath $x$},\tau)= 
\int\frac{d^3k}{(2\pi)^3}\,{\cal X}_\mu^\alpha
(\mbox{\boldmath\bf $k$},\tau) 
\exp(i\mbox{\boldmath\bf $k\cdot x$})
\label{fourX}
\end{equation}

Recalling that \mbox{$\varphi=M+\varphi_0P_\varphi(\omega_\varphi\tau)$} and 
using (\ref{masses}) we can find the effective mass ${\cal M}_{XY}$
of $X_\mu^\alpha$ during preheating to be,

\begin{equation}
M_{XY}^2=\frac{1}{2}g^2(M^2+\varphi_0^2P_\varphi^2+2M\varphi_0P_\varphi)
\Rightarrow 
{\cal M}_{XY}^2
=\frac{1}{2}g^2M^2
\Big[1+\frac{1}{2}(\frac{\varphi_0}{M})^2\Big]\simeq\frac{3}{4}g^2M^2
\label{MXYeff}
\end{equation}
where we used that \mbox{$\langle P_\varphi^2\rangle_\tau =1/2$} and 
\mbox{$\langle P_\varphi\rangle_\tau = 0$}. 
Rescaling the time as 
\mbox{$\omega_\varphi\tau\rightarrow\varpi$}, the equations
(\ref{Xparal}) and (\ref{Xperp}) can be recast as,

\begin{eqnarray}
\left[\Big(\partial_\varpi^2+
\frac{k^2+M_{XY}^2}{\omega_\varphi^2}\Big)+
\frac{2(\partial_\varpi\ln M_{XY})k^2}{k^2+M_{XY}^2}\partial_\varpi
\right]
\mbox{\boldmath $\cal X$}^{\alpha\parallel}
 & = & 0\label{Xparal1}\\
 & & \nonumber\\
\Big(\partial_\varpi^2+
\frac{k^2+M_{XY}^2}{\omega_\varphi^2}\Big)
\mbox{\boldmath $\cal X$}^{\alpha\perp}
 & = & 0\label{Xperp1}
\end{eqnarray}

Now, it easy to show that, in the above, we have,

\begin{equation}
\frac{k^2+M_{XY}^2}{\omega_\varphi^2}=
\frac{k^2+{\cal M}_{XY}^2}{\omega_\varphi^2}+\frac{g^2M^2}{2\omega_\varphi^2}
\Big[\frac{1}{2}(\frac{\varphi_0}{M})^2\hat{P}_\varphi+
2\frac{\varphi_0}{M}P_\varphi\Big]
\end{equation}
where \mbox{$\hat{P}_\varphi=2P_\varphi^2-1$} is a periodic function of 
$\varpi$ with frequency \mbox{$\hat{\omega}_\varphi=2\omega_\varphi$} and 
max\{$|\hat{P}_\varphi|$\}=1. 
Each of the above equations contains two resonant channels with the 
$q$-factors,

\begin{eqnarray}
q_1 & = & \frac{g^2M^2}{2\omega_\varphi^2}\frac{\varphi_0}{M} 
\simeq \frac{g^2}{8\lambda}
\qquad\nonumber\\
q_2 & = & \frac{g^2M^2}{8\omega_\varphi^2}(\frac{\varphi_0}{M})^2
\simeq\frac{g^2}{32\lambda}\simeq\frac{1}{4}q_1
\nonumber\\
A_{k=0} & = & \frac{{\cal M}_{XY}^2}{\omega_\varphi^2}
\simeq \frac{3g^2}{16\lambda}\simeq\frac{3}{2}q_1\simeq 6q_2
 \label{XqqA}
\end{eqnarray}

This implies that, provided $q_{1,2}\gsim 1/2$, 
the $q_1$-resonance is efficient in amplifying gauge
fields, while the $q_2$-resonance is inefficient.
In fact when $A_{k=0}<2q$ \{$A_{k=0}>2q$\} the growth rate of the field
grows \{decreases\} with increasing $q$~\cite{charts1}, implying that 
for \mbox{$A_{k=0}\simeq\frac{3}{2}q_1$} the resonance becomes
stronger as $q_1$ increases. For example, one enters the first broad
resonance at about $q_1\simeq 1/2$ and exits it at about $q_1\simeq 3$. 

When compared with the transverse equation~(\ref{Xperp1}), 
the longitudinal equation~(\ref{Xparal1}) contains an  
additional `damping' term. The sign of the `damping' coefficient 
\mbox{$\partial_\varpi\ln M_{XY}=\partial_\varpi P_\varphi/P_\varphi$}
may be either positive or negative, resulting in damping or 
growth, respectively. Since the positive and negative parts are symmetrically
and evenly distributed, we expect that the growth wins over the
damping, just as it is in the case of stochastic resonance \cite{stocha}. 
As a consequence, the longitudinal modes grow faster than
the transverse modes. Large longitudinal amplitudes can be mediated
to the transverse modes through scatterings. We do not consider here 
possible physical consequences of these processes. 

Similar results are obtained for the $Y^\alpha_\mu$ and the $V_\mu$ bosons, 
substituting, in the latter case \mbox{$M_{XY}\rightarrow M_V$}
or equivalently \mbox{$g\rightarrow 2\sqrt{(6/5)}\,g/\cos\Theta$}.

\subsubsection{The amplitudes of the massive vector bosons}

In order for the backreaction to stop resonant production of the
massive gauge fields, the induced (Hartree) mass must be larger than
about the mass of the Higgs boson, since this is the typical resonance
scale. This can be argued as follows. The shift in $A$ 
required to switch off the leading resonance is of the order
$\delta A\sim 0.5$. This cannot be mediated by the $\varphi$-field
modes, since the maximum possible expectation value is
\mbox{$\langle(\delta \varphi)^2\rangle_{\rm max}\sim M^2$}, 
resulting in 
\mbox{$\delta A=\delta {\cal M}_{XY}^2/\omega_\varphi^2\sim g^2/16\lambda$},
which is typically not sufficient to shut down the resonance. 
Moreover, the inflaton does not help, since it does not couple to
the vector bosons. 

There are however cubic terms in the source current, which have been
neglected in~(\ref{Xperp1}) and which may shut down the 
resonance. The cubic source current terms can be obtained from (\ref{J}).
Similarily to (\ref{cubic}), for the supermassive GUT-bosons we have,

\begin{equation}
(J^\nu_{\xi\alpha})^{\rm cubic}_{XY}=g^2g^{\mu\rho}g^{\nu\sigma}
Re[(\w_\mu)_{\xi i}(\w_\rho)_{ij}(\w_\sigma)_{j\alpha}-
(\rho\leftrightarrow\sigma)]
\label{oneJ}
\end{equation}
where \mbox{$\xi=4,5$} and \mbox{$\alpha=1,2,3$}. In view of (\ref{W}),
(\ref{oneJ}) can be written schematically as 
\footnote{Here we have used that 
\mbox{$\frac{1}{c_V}(\frac{1}{c_V}+c'_V)=1$}.}

\begin{equation}
J^{\rm cubic}_{XY}\sim g^2(XXX+WWX+WVX+VVX),
\end{equation} 
where $X$ denotes the supermassive GUT-bosons $X^\alpha_\mu$ and 
$Y^\alpha_\mu$, $W$ denotes the electroweak
gauge bosons $W^\pm_\mu,\,Y_\mu$, which are gravitationally produced in 
inflation as shown in Sec. \ref{mech}, and $V$ is the massive 
$V$-boson. Note that the contribution from the $G$-bosons of SU(3)$_c$
is negligible because they are neither produced gravitationally 
in inflation, nor are they generated during preheating, since they do
not couple to any Higgs field. Thus, a typical contribution of the
backreaction current to the mass of the $X^\alpha_\mu, Y^\alpha_\mu$
fields, coming from their resonant production, is of the form

\begin{equation}
\delta M^2_{XY}\sim  g^2(\langle X^2\rangle+\langle V^2\rangle)
\end{equation}

Similarly, for $V_\mu$ we have from (\ref{feV}) that 
$J^\nu_V=-\sqrt{\frac{5}{12}}\,\cos\Theta J_Y^\nu$ so that the cubic term
contribution is given by (\ref{JYV}) to be,

\begin{equation}
J^{\rm cubic}_V\sim \frac{5}{6}\,g^2XXV
\end{equation} 

Hence, the contribution of the backreaction current to the mass of the 
$V_\mu$ is,

\begin{equation}
\delta M^2_V\sim \frac{5}{6}\,g^2\langle X^2\rangle
\end{equation}

In order to get the first resonance to shut
down, one requires $\delta M^2\sim M_H^2 = 2\lambda M^2$.
Hence, the energy stored in the $X$ and $V$ fields is then of the order,

\begin{equation}
\rho_{XV}\sim g^2(\langle X^2\rangle^2+\langle V^2\rangle^2)
\sim\frac{\lambda^2}{g^2}M^4
\sim\frac{\lambda}{g^2}\rho_{\rm inf}
\end{equation}

Since \mbox{$\lambda\sim 1$} we conclude 
that the resonance shuts down only when most of the energy is in the massive 
gauge field modes. In other words, the resonance is efficient until a 
significant fraction of the inflaton's energy decays. Then, as the oscillating 
amplitude of $\varphi$ decreases, one enters the narrow resonance and the 
decay slows down. This is just like the case of two scalar fields with 
$q\sim 1$, and also the one field case, where also $q\sim 1$. One is then 
left with the condensates of the gauge fields whose amplitude can be estimated 
to be,

\begin{equation}
|X^\alpha_\mu|,|Y^\alpha_\mu|, |V_\mu| 
\sim g_*^{-1/4}\frac{M_H}{g} \sim 0.1 M_H
\label{bound}
\end{equation}
where \mbox{$g_*\lsim 100$} is the effective number of degrees of freedom
produced by the resonance. 

\subsection{The hypercharge field at preheating}

In a similar manner as for the supermassive GUT-bosons one finds from 
(\ref{feY}) that the hypercharge transverse mode equation is,

\begin{equation}
\Big(\partial_\tau^2+k^2\Big)\mbox{\boldmath $\cal Y$}^{\perp}=
\sqrt{\mbox{$\frac{5}{12}$}}\sin\Theta 
\mbox{\boldmath $\cal J$}_{\!Y}^{\perp}
\label{Yperp}
\end{equation}
where 

\begin{eqnarray}
J_Y^\nu(\mbox{\boldmath $x$},\tau)\!=\!\! 
\int\!\frac{d^3k}{(2\pi)^3}\,{\cal J}_{\!Y}^\nu
(\mbox{\boldmath\bf $k$},\tau) 
\exp(i\mbox{\boldmath\bf $k\cdot x$}) & \mbox{and} & 
Y^\nu(\mbox{\boldmath $x$},\tau)\!=\!\! 
\int\!\frac{d^3k}{(2\pi)^3}\,{\cal Y}^\nu
(\mbox{\boldmath\bf $k$},\tau) 
\exp(i\mbox{\boldmath\bf $k\cdot x$}) 
\end{eqnarray}
and we have used the current conservation equation, 
\mbox{$\partial_\tau({\cal J}_Y)_\tau=i\mbox{\boldmath $k\cdot\cal J$}_{\!Y}$}.

Using the Hartree approximation, as shown in the appendix \ref{a1}, and also in
view of (\ref{JYbig}) we can recast (\ref{Yperp}) as, 

\begin{equation}
\left\{\partial_\tau^2 + k^2
+6\,\bar{g}_Y^2
\Big[\mbox{\normalsize $\frac{2}{3}$}
\langle|\mbox{\boldmath $X$}|^2\rangle_{_V}
-\langle |X_\tau|^2\rangle_{_V}\Big]
\right\}\mbox{\boldmath ${\cal Y}$}^{\perp}= 
-6\,\bar{g}_Y^2\cot\Theta 
\Big[
\mbox{\normalsize $\frac{2}{3}$}\langle|\mbox{\boldmath $X$}|^2\rangle_{_V}
-\langle |X_\tau|^2\rangle_{_V}\Big]\mbox{\boldmath ${\cal V}$}^{\perp}
\label{Yres}
\end{equation}
where \mbox{$\bar{g}_Y^2\equiv\frac{5}{6}g_Y^2$} and $X_\mu$ stands for
all the 6 the supermassive GUT-bosons $X_\mu^\alpha,Y_\mu^\alpha$ with,

\begin{eqnarray}
\langle|\mbox{\boldmath $X$}|^2\rangle_{_V}
 & = &
\frac{1}{V}\int \frac{k^2d k}{2\pi^2}
\mbox{\boldmath ${\cal X}$}(\mbox{\boldmath $k$},\tau)
\mbox{\boldmath ${\cal X}$}(-\mbox{\boldmath $k$},\tau)
\;\;\;\,\simeq \;\;
\frac{1}{V}\int \frac{k^2d k}{2\pi^2}
|\mbox{\boldmath ${\cal X}$}(k,\tau)|^2\\
 & & \nonumber\\
\langle|X_\tau|^2\rangle_{_V}
 & = & \frac{1}{V}\int \frac{k^2d k}{2\pi^2}
{\cal X}_\tau(\mbox{\boldmath $k$},\tau)
{\cal X}_\tau(-\mbox{\boldmath $k$},\tau)
\;\;\simeq\;\;
\frac{1}{V}\int \frac{k^2d k}{2\pi^2}
[{\cal X}_\tau(k,\tau)]^2
\nonumber
\end{eqnarray}
and also [{\em cf.} (\ref{calZ1}) and (\ref{longtrans})],

\begin{equation}
X_{\rm rms}^2 \equiv\frac{2}{3}\langle|\mbox{\boldmath $X$}|^2\rangle_{_V}
-\langle |X_\tau|^2\rangle_{_V}
=\frac{1}{V}\int\frac{k^2dk}{2\pi^2}\left\{
\frac{2}{3}[\mbox{\boldmath ${\cal X}$}(\tau)]^2
+\frac{k^2[\partial_\tau 
\mbox{\boldmath ${\cal X}$}^{\parallel}(\tau)]^2}{k^2+M_{XY}^2}\right\}
\end{equation}

We shall now investigate whether the hypercharge field~(\ref{Yres})
may undergo resonant amplification at preheating. Since the supermassive
GUT-bosons are mainly produced at preheating, the dominant contribution to 
$X_{\rm rms}^2$ comes from the resonant modes that initially oscillate in 
phase~\cite{charts1}, which in principle may drive resonant amplification of 
the hypercharge. We now discuss the necessary conditions required for this 
indirect resonance to be operative.

At early times in preheating the Hartree terms $X_{\rm rms}^2$ are small,
so that the hypercharge bosons grow through (inefficient) narrow parametric 
resonance~\cite{branden2}.
When the oscillatory component of $X_{\rm rms}^2$ becomes of the order
$\omega_\varphi^2$ however, the hypercharge resonance can become 
broad~\cite{charts2}. 
Since the hypercharge field has an infrared spectrum that is already
amplified in inflation, while the $V$-boson spectrum is that of the Minkowski
vacuum, when compared with the $X^2V$-term, the $X^2Y$-term is expected
to dominate the resonant growth on superhorizon scales. 
Therefore, we may write, 

\begin{equation}
\mbox{\boldmath $\cal J$}_{\!Y}^{\perp}(\tau)\simeq
-6\sqrt{2}\,g\bar{g}_YX_{\rm rms}^2
\mbox{\boldmath ${\cal Y}$}^{\perp}(\tau)
\label{YXXY}
\end{equation}
so that (\ref{Yres}) simplifies to 

\begin{equation}
\frac{d^2\mbox{\boldmath $\cal Y$}^\perp}{d\tau^2}
+\left( k^2 + 6\,\bar{g}_Y^2 X_{\rm rms}^2 \right)
\mbox{\boldmath $\cal Y$}^\perp\simeq 0
\label{Ydynamo}
\end{equation}

Now, $X_{\rm rms}^2$ can be conveniently split into
the slowly growing and oscillatory contributions as follows: 

\begin{equation}
 X_{\rm rms}^2 = X_0^2(\tau) + X_{\rm osc}^2 [1+ P_X(\tau)]
\label{Xvariance}
\end{equation}
where $P_X(\tau)$ is a periodic function with 
period \mbox{$\sim\pi/\omega_\varphi$} such that 
\mbox{$\langle P_X(\tau)\rangle_\tau=0$} and 
\mbox{$\langle P_X^2(\tau)\rangle_\tau=1/2$}, and 

\begin{equation}
X_0^2(\tau) = X_{\rm vac}^2 \left(e^{2\mu_0\omega_\varphi \tau} -1\right)
 ,\qquad
X_{\rm osc}^2(\tau) 
=X_{\rm vac}^2\left(e^{2\mu_X\omega_\varphi \tau} -1\right)
\label{Xvariance2}
\end{equation}
where $\mu_X$ denotes the relevant resonant growth rate
of the supermassive GUT-boson fields $X$, and $\mu_0$ is the growth rate of 
the constant contribution to the $X$-field mass induced by the 
inelastic scatterings. Also, we have subtracted the vacuum contribution 
\mbox{$X_{\rm vac}^2\sim k_{\rm res}^2$}, with
$k_{\rm res}^2$ being the typical resonant momentum.
Provided \mbox{$\mu_0<\mu_X$}, or equivalently 
\mbox{$X_{0}<X_{\rm osc}$}, the hypercharge grows resonantly. 
The amount of available energy \mbox{$\rho_{\rm inf}\simeq\lambda M^4/4$}
results in the following upper bound,

\begin{equation}
\frac{1}{2} {\cal M}_{XY}^2 X_{\rm rms}^2 \lsim \frac{1}{8} M_H^2 M^2
\label{Xvariance3}
\end{equation}
or equivalently [{\em cf.} (\ref{MXYeff})]

\begin{equation}
X_{\rm rms}^2 \lsim \frac{2\lambda}{3g^2} M^2
\label{Xvariance4}
\end{equation}
where \mbox{${\cal M}_{XY}^2\simeq 3g^2M^2/4$} and 
\mbox{$M_H^2=2\lambda M^2$}. This is in agreement with (\ref{bound}) and
implies the following bound for the quality factor of the hypercharge 
resonance,

\begin{equation}
q_Y = \frac{6\,\bar g_Y^2  X_{\rm osc}^2}{2\omega_\varphi^2}
     \lsim \frac{5}{12}\sin^2\Theta<0.4
\label{Xvariance5}
\end{equation}
where \mbox{$\sin\Theta=g_Y/g$}. Since significant resonant amplification
of the hypercharge on superhorizon scales may occur without tuning only when 
$q_Y\gsim 1/2$, we conclude that {\em the inflationary spectrum of the 
hypercharge gauge bosons is hardly amplified by parametric resonance.} 

\section{Conclusions}

We have presented a mechanism of primordial magnetic field generation based
on the breaking of conformal invariance of the $Z$-boson field during 
inflation. The mechanism is generic and independent of inflationary model
as long as the reheating temperature is higher than the electroweak scale.
This is because our mechanism requires a phase of electroweak unification to 
``channel'' the generated superhorizon $Z$-boson spectrum into the
photon, through the hypercharge field. 

The conformal invariance of $Z$ is
naturally broken due to its standard-model coupling with the electroweak
Higgs-field. The latter, during inflation,
develops a condensate comparable to the one of the inflaton field itself 
resulting in a non-zero, but small, mass for the $Z$-boson field. 
Conformal invariance breakdown results in gravitational production of the
$Z$-field on superhorizon scales. We have computed the relevant superhorizon 
spectrum and found that it is almost scale invariant. At the end of inflation
reheating restores the electroweak symmetry and the $Z$-spectrum is converted 
into a hypercharge spectrum (the photon's contribution being negligible), 
which, due to the stochastic nature of the original $Z$-fluctuations, gives 
rise to a superhorizon hypermagnetic field that freezes 
into the reheated plasma. We have calculated the spectrum of the rms value
of the hypermagnetic field and found that it is of the form 
\mbox{$B_{\rm rms}\propto 1/\ell$} as shown in Fig. \ref{results}. 
In a similar way non-Abelian $W$-fields are also 
generated but their associated magnetic fields are screened due to the 
existence of a magnetic mass of non-perturbative nature, which stems 
from their self-interaction. During the radiation era the hypermagnetic field 
evolves satisfying flux conservation. At the electroweak phase transition 
it is transformed into a regular magnetic field. When scaled until 
galaxy formation, this magnetic field is found to be sufficient to trigger the 
galactic dynamo and explain the observed galactic magnetic fields in the case 
of a spatially-flat, dark-energy dominated Universe with GUT-scale inflation. 

The beauty of our mechanism, apart from being independent of the inflationary 
model, lies in that it does not, unlike most proposed mechanisms, require
the explicit introduction of conformal invariance breaking terms in 
the Lagrangian or any exotic fields other than the ones of the standard 
model and the inflaton field. Moreover, we do not have to specify 
a particular GUT group, or involve grand unification in any way. 

An intriguing possibility was that our magnetic field could be further 
amplified during preheating. In order to study this we considered a 
(supersymmetric) hybrid inflationary model and grand unification under
Flipped~SU(5), both well motivated. However, we have found that preheating 
amplification is probably negligible since the hypercharge field is amplified 
only via indirect narrow resonance which turns out to be rather inefficient. 
Still, a better (possibly numerical) treatment of the magnetic field at 
preheating is probably necessary to provide an adequate understanding of its 
behavior. Such treatment should, in principle, address issues such as the 
conductivity of the newly created plasma or the fate of the hyperelectric 
field. 

In particular, the behavior of the conductivity during preheating may 
inhibit our hypermagnetic field. However, an analytic 
treatment of this issue is rather complex and beyond the scope of the present 
article and numerical studies are inconclusive. 
It seems that the behavior of the conductivity (spatial distribution 
and growth rate) is highly non-trivial, non-perturbative and model dependent 
\cite{bassett}. It has been shown however, that the appearance of conductivity 
in preheating does not necessarily inhibit the growth or the existence of 
(hyper)magnetic fields but may allow substantial amplification 
\cite{bassett}\cite{finelli}. This is certainly an issue that deserves further 
investigation.

In summary, we have presented a natural mechanism for magnetogenesis
by inflation which may be an explanation for galactic magnetic fields.

\bigskip

\noindent
{\Large\bf Acknowledgements}

\bigskip

\noindent
Support for K.D. was provided by DGICYT grant PB98-0693 and by the
European Union under contract HPRN-CT00-00148; 
O.T. wishes to thank the Theoretical Physics Group at
Imperial College for a Visiting Fellowship.
We gratefully acknowledge travel support
by the U.K. PPARC. This work is supported in part by PPARC.

\appendix

\section{The Hartree approximation}\label{appa}

In this appendix we discuss the use of the Hartree approximation in the
computation of scalar and gauge fields. 
This approximation is applicable in this case mainly due to the 
classicallity of the superhorizon modes and also
since the typical value of the quality factor $q$ of the parametric 
resonance is about one or smaller, 
essentially until the field grows to its maximum value.\footnote{The 
validity of the Hartree approximation can be checked by performing a 
consistent one-loop calculation.}

The Hartree approximation takes account of elastic scatterings only, which do 
not change the momenta of incoming particles, and it models the dynamics well 
when 2-to-2 scatterings are the only ones that are relevant. This is indeed so
at the early stages of preheating, as is known from exact classical 
simulations (e.g. of the scalar theory). When 
thermalization begins, inelastic scatterings, which include 2-to-4, etc., 
become important. They combine the infrared modes into more energetic ones, as 
it is required by the thermalization process, because particles produced by 
parametric resonance are more infrared in comparison to a thermal 
distribution. The Hartree approximation breaks down after a significant 
fraction of the inflaton's energy has decayed.

The Hartree approximation consists essentially in replacing a product of two 
fields (in an equation of motion) with the spatial average as follows,

\begin{equation}
A(\mbox{\boldmath $x$},t)\,
B(\mbox{\boldmath $x$},t)\rightarrow 
\langle 
A(\mbox{\boldmath $x$},t)\,
B(\mbox{\boldmath $x$},t)
\rangle_{_V}=\tilde{A}\tilde{B}
+\langle\delta A(\mbox{\boldmath $x$},t)\,
\delta B(\mbox{\boldmath $x$},t)
\rangle_{_V}
\end{equation}
where $\tilde{A}$ and $\tilde{B}$ denote the zero modes defined as
\mbox{$\tilde{A}\equiv\langle A(\mbox{\boldmath $x$},t)\rangle_{_V}$},
\mbox{$\tilde{B}\equiv\langle B(\mbox{\boldmath $x$},t)\rangle_{_V}$}
and also\\
\mbox{$\delta A(\mbox{\boldmath $x$},t)=A(\mbox{\boldmath $x$},t)-\tilde{A}$},
\mbox{$\delta B(\mbox{\boldmath $x$},t)=B(\mbox{\boldmath $x$},t)-\tilde{B}$}, 
with the volume averages being defined as,

\begin{eqnarray}
\langle
\delta A(\mbox{\boldmath $x$},t)\,
\delta B(\mbox{\boldmath $x$},t)
\rangle_{_V} 
 & \equiv & \frac{1}{V}\int d^3x\,
\delta A(\mbox{\boldmath $x$},t)\,\delta B(\mbox{\boldmath $x$},t)
\;=\;\frac{1}{V}\int^\prime \frac{d^3 k}{(2\pi)^3}
{\cal A}(\mbox{\boldmath $k$},t)\,{\cal B}(-\mbox{\boldmath $k$},t)
\nonumber\\
\tilde{A} & \equiv & 
\frac{1}{V}\int d^3x \, A(\mbox{\boldmath $x$},t)
=\frac{1}{V}\;{\cal A}(\mbox{\boldmath $k$}\!=\!0,t)
\end{eqnarray}

In these definitions the volume $V$ is defined by the following 
discretized version of the momentum integral

\begin{equation}
\int \frac{d^3 k}{(2\pi)^3}\rightarrow\frac{1}{V}
\sum_{\mbox{\scriptsize\boldmath $k$}}
\end{equation}

The prime in the momentum integral indicates that the zero mode should be
taken out of the integral. The zero modes require special care, since they 
may become macroscopic, leading to a condensate.\footnote{Note that this 
ensemble averaging corresponds to the classical ensemble representation of 
the quantum state.}
 
\subsection{Hypercharge source current}\label{a1}

Let us employ the Hartree approximation in computing the source current of
the hypercharge which is shown explicitly in (\ref{JYbig}).
In the spirit of our approximation, 
we shall neglect the terms in which the zero modes figure incoherently in
quadratic or cubic combinations. This means that at this point we shall
take account only of those nonlinear contributions from the zero modes which 
oscillate coherently in time.

Some of the remaining terms are composed of the fields which have incoherent
phases in different points in space, so that, when averaged over space, 
they vanish. Including these terms is strictly speaking beyond the Hartree 
approximation.\footnote{One way of improving on the Hartree approximation is 
to treat the terms that incoherently contribute as noise. 
When averaged over time, these terms yield zero. In \cite{branden} it has been 
shown that including these noise terms can only enhance resonant production. 
Thus, when not considering these terms we adopt a conservative approach on the 
resonant growth.} Therefore, to first approximation we keep only 
the terms that contribute to the genuine Hartree approximation,
which are the terms that oscillate coherently in space and time, since they 
dominate the resonant production.

Consider first the derivative terms of the source current in 
(\ref{JYbig}),

\begin{eqnarray}
-\mbox{\normalsize $\sqrt{\frac{5}{12}}$}
\,\sin\Theta\,J_Y^\nu[X\partial X] & \equiv &
-\;\bar{g}_Y\,
\mbox{Im}\Big\{[\partial_\mu+(\partial_\mu\ln\sqrt{-D_g})]
g^{\mu\rho}g^{\nu\sigma}
(X^\alpha_\rho\overline{X}^\alpha_\sigma+
Y^\alpha_\rho\overline{Y}^\alpha_\sigma)\,+\nonumber\\
 & & \\
 & & +\;g^{\mu\rho}g^{\nu\sigma}
[X^\alpha_\mu\partial_\rho\overline{X}^\alpha_\sigma
+Y^\alpha_\mu\partial_\rho\overline{Y}^\alpha_\sigma
-(\rho\leftrightarrow\sigma)]\Big\}\nonumber
\end{eqnarray}

In CFRW with \mbox{$a\simeq 1$} the above may be written as,

\begin{equation}
-\mbox{\normalsize $\sqrt{\frac{5}{12}}$}
\sin\Theta
J^Y_\rho[X\partial X]=\!-\bar{g}_Y\eta^{\mu\sigma}
\mbox{Im}\Big\{
\partial_\mu(X^\alpha_\rho\overline{X}^\alpha_\sigma+
Y^\alpha_\rho\overline{Y}^\alpha_\sigma)+
X^\alpha_\mu(\partial_\rho\overline{X}^\alpha_\sigma-
\partial_\sigma\overline{X}^\alpha_\rho)
+Y^\alpha_\mu(\partial_\rho\overline{Y}^\alpha_\sigma-
\partial_\sigma\overline{Y}^\alpha_\rho)\Big\}
\label{JXdX}
\end{equation}

Hartree averaged the terms in the above are of the form,

\begin{eqnarray}
\langle X^\alpha_\mu\partial_0
\overline{X}^\alpha_\sigma\rangle_{_V}\equiv
\frac{1}{V}\int d^3\!x\,
X^\alpha_\mu\partial_0\overline{X}^\alpha_\sigma & = &
\frac{1}{V}\int\frac{d^3k}{(2\pi)^3}\,
{\cal X}^\alpha_\mu(\mbox{\boldmath $k$},\tau) \,
\partial_\tau\overline{{\cal X}}^\alpha_\sigma
(\mbox{\boldmath $k$},\tau)\nonumber\\
 & & \\
 \langle X^\alpha_\mu\partial_i 
\overline{X}^\alpha_\sigma\rangle_{_V}\equiv
\frac{1}{V}\int d^3\!x\,
X^\alpha_\mu\partial_i\overline{X}^\alpha_\sigma & = &
\frac{1}{V}\int\frac{d^3k}{(2\pi)^3}\,
{\cal X}^\alpha_\mu(\mbox{\boldmath $k$},\tau)\, 
ik_i\,\overline{{\cal X}}^\alpha_\sigma(\mbox{\boldmath $k$},\tau)\nonumber
\end{eqnarray}
where ${\cal X}^\alpha_\mu$ is defined in (\ref{fourX}). 
As a consequence of temporal and spatial isotropy, we have
\mbox{$\langle X^\alpha_\mu\partial_\rho 
\overline{X}^\alpha_\sigma\rangle_{_V}=
\tilde{X}^\alpha_\mu\partial_\rho\tilde{\overline{X}}^\alpha_\sigma
\propto \eta_{\mu\sigma}$} that is,

\begin{equation}
\langle X^\alpha_\mu\partial_\rho 
\overline{X}^\alpha_\sigma\rangle_{_V}=
\frac{1}{2}\eta_{\mu\sigma}\partial_\rho\sum_\alpha|\tilde{X}^\alpha_\sigma|^2
\Rightarrow
\mbox{Im}\Big(\langle X^\alpha_\mu\partial_\rho 
\overline{X}^\alpha_\sigma\rangle_{_V}\Big)=0
\end{equation}
where there is no summation over the $\sigma$ index. 
Since all the terms in (\ref{JXdX}) are of
the same form we conclude that, in the Hartree approximation, 
\mbox{$J^Y_\rho[X\partial X]=0$}, i.e. 
the spatial averages of the quadratic derivative terms in
(\ref{JYbig}) average to zero and only the cubic terms survive.

Let us consider the cubic terms now. We start by noting that, for two
{\em different} gauge fields $A_\mu$ and $B_\mu$ we have, 
\mbox{$\langle A_\mu B_\nu\rangle_{_V}=\eta_{\mu\nu}
\tilde{A}_\rho\tilde{B}^\rho$}.
The zero modes are proportional to the polarization vectors, that is, 

\begin{eqnarray}
\tilde{A}_\rho^{p}=\frac{1}{V}{\cal A}^p_\rho(\mbox{\boldmath $k$}\!=\!0,\tau)
\propto \epsilon^p_{\rho\,\mbox{\scriptsize\boldmath $k$}\rightarrow 0}
 & \mbox{and} & 
\tilde{B}^{p'}_\rho=\frac{1}{V}{\cal B}^{p'}_\rho
(\mbox{\boldmath $k$}'\!=\!0,\tau)\propto 
\epsilon^{p'}_{\rho\,\mbox{\scriptsize\boldmath $k$}^\prime\rightarrow 0} ,
\end{eqnarray}
where $p=T,L$ ($T=1,2$ and $L=3$) denotes polarization, and 
\mbox{$\epsilon^p_{\rho\,\mbox{\scriptsize\boldmath $k$}} 
(\epsilon^{p'}_{\mbox{\scriptsize\boldmath $k$}'})^\rho
=-\mbox{\boldmath $\epsilon$}^p_{\mbox{\scriptsize\boldmath $k$}}\cdot 
\mbox{\boldmath $\epsilon$}^{p'}_{\mbox{\scriptsize\boldmath $k$}'}
=-\delta_{\mbox{\scriptsize\boldmath $k$},\mbox{\scriptsize\boldmath $k$}'}\,
\delta^{pp'}$}.

As a consequence of the vectorial nature of the condensates,
and their origin in quantum fluctuations, we conclude that they 
are randomly oriented, and hence, for any different gauge fields,

\begin{equation}
\langle A_\mu B^\nu\rangle_{_V}
= \eta_{\mu\nu}\tilde{A}_\rho \tilde{B}^\rho\equiv 0
\end{equation}

In view of the above result the hypercharge source current in (\ref{JYbig})
in the Hartree approximation can be recast as,

\begin{equation}
\begin{array}{ccl}
-\sqrt{\frac{5}{12}}\,\sin\Theta\,J_Y^\nu & = &
-\;\bar{g}^2_Y{\cal E}^{\mu\nu\rho\sigma}
\mbox{Re}\Big(\langle X^\alpha_\mu\overline{X}^\alpha_\sigma\rangle_{_V}
+\langle Y^\alpha_\mu\overline{Y}^\alpha_\sigma\rangle_{_V}\Big)\,Y_\rho-\\
 & & \\
 & & -\;\cot\Theta\,\bar{g}_Y^2{\cal E}^{\mu\nu\rho\sigma}
\mbox{Re}\Big(\langle X^\alpha_\mu\overline{X}^\alpha_\sigma\rangle_{_V}
+\langle Y^\alpha_\mu\overline{Y}^\alpha_\sigma\rangle_{_V}\Big)\,V_\rho-\\
 & & \\
 & & -\;\frac{1}{\sqrt{2}}\,\bar{g}_Yg\,{\cal E}^{\mu\nu\rho\sigma}
\mbox{Re}\Big(\langle Y^\alpha_\mu\overline{Y}^\alpha_\sigma\rangle_{_V}-
\langle X^\alpha_\mu\overline{X}^\alpha_\sigma\rangle_{_V}\Big)W^3_\rho
\end{array}
\end{equation}

Because the $X^\alpha_\mu$ and $Y^\alpha_\mu$ bosons are 
entirely equivalent we define, 

\begin{equation}
\langle X_\mu\overline{X}_\sigma\rangle_{_V}\equiv
\langle X^\alpha_\mu\overline{X}^\alpha_\sigma\rangle_{_V}=
\langle Y^\alpha_\mu\overline{Y}^\alpha_\sigma\rangle_{_V}
\hspace{1cm}\forall\;\alpha =1,2,3\hspace{1cm}(\mbox{no summation})
\end{equation}

Thus, the above becomes,

\begin{equation}
-\mbox{\normalsize $\sqrt{\frac{5}{12}}$}
\,\sin\Theta\,
J_Y^\nu=-6\,\bar{g}^2_Y{\cal E}^{\mu\nu\rho\sigma}
\mbox{Re}\Big(\langle X_\mu\overline{X}_\sigma\rangle_{_V}\Big)\,Y_\rho-
6\,\cot\Theta\,\bar{g}_Y^2{\cal E}^{\mu\nu\rho\sigma}
\mbox{Re}\Big(\langle X_\mu\overline{X}_\sigma\rangle_{_V}\Big)\,V_\rho
\end{equation}

Therefore, in the Hartree approximation only the terms of the $Y_\mu$ and 
$V_\mu$ bosons contribute to the hypercharge source current. In view of the
definition (\ref{BigE}) it is straightforward to show that, 

\begin{equation}
{\cal E}^{\mu\nu\rho\sigma}
\mbox{Re}\Big(\langle X_\mu\overline{X}_\sigma\rangle_{_V}\Big)=
\mbox{Re}\langle X^\rho\overline{X}^\nu\rangle_{_V}-\eta^{\rho\nu}
\langle X_\sigma\overline{X}^\sigma\rangle_{_V}
\end{equation}

Thus the source current becomes,

\begin{equation}
-\mbox{\normalsize $\sqrt{\frac{5}{12}}$}
\,\sin\Theta\,
J_Y^\nu =-6\,\bar{g}^2_Y
\Big(\mbox{Re}\langle X^\rho\overline{X}^\nu\rangle_{_V}-\eta^{\rho\nu}
\langle X_\sigma\overline{X}^\sigma\rangle_{_V}\Big)(Y_\rho+\cot\Theta\,V_\rho)
\end{equation}

In (\ref{Yres}) we need the spatial component of the above, for which we find,

\begin{equation}
-\mbox{\normalsize $\sqrt{\frac{5}{12}}$}\,\sin\Theta\,
(J_Y)_i=6\,\bar{g}^2_Y\Big[\langle|X_0|^2\rangle_{_V}-
\frac{2}{3}\langle|\mbox{\boldmath $X$}|^2\rangle_{_V}\Big]
(Y_i+\cot\Theta\,V_i)
\end{equation}
where \mbox{$|X_0|^2\equiv X_0\overline{X}_0$},
\mbox{$|\mbox{\boldmath $X$}|^2\equiv X_j\overline{X}_j$} and we used
\mbox{$\langle X_1\overline{X}_1\rangle_{_V}=
\langle X_2\overline{X}_2\rangle_{_V}=
\langle X_3\overline{X}_3\rangle_{_V}=\frac{1}{3}
\langle|\mbox{\boldmath $X$}|^2\rangle_{_V}$}. Because  
$\langle\cdot\rangle_{_V}$ is a function of time only, we can immediately
obtain the Fourier transformed current, 

\begin{equation}
-\mbox{\normalsize $\sqrt{\frac{5}{12}}$}
\,\sin\Theta\,\mbox{\boldmath ${\cal J}$}_Y^\perp=
6\,\bar{g}^2_Y\Big[\langle|X_\tau|^2\rangle_{_V}-
\frac{2}{3}\langle|\mbox{\boldmath $X$}|^2\rangle_{_V}\Big]
(\mbox{\boldmath ${\cal Y}$}^\perp+
\cot\Theta\,\mbox{\boldmath ${\cal V}$}^\perp)
\end{equation}
which is used in (\ref{Yres}).

\subsection{Gravitational production during inflation}

Here we will briefly calculate the volume averages of scalar and gauge fields
which are gravitationally produced during inflation. In particular we will 
focus on the $Z$-boson field and the electroweak Higgs field $\psi$.

\subsubsection{\boldmath The case of the $Z$ boson}\label{a21}

For the $Z$-boson field we have,

\begin{equation}
\langle Z^2\rangle_{_V}\equiv 
\langle 0\vert\mbox{\boldmath ${\cal Z}\cdot{\cal Z}$}\vert 0\rangle
-\langle 0\vert(\mbox{\boldmath ${\cal Z}\cdot{\cal Z}$})_{\rm vac}
\vert 0\rangle
=\frac{1}{V}\int\!\!\frac{d^3k}{(2\pi)^3}\, 
\sum_p\left[{\cal Z}^{(1)}_p (k\tau){\cal Z}^{(2)}_p(k\tau)
-\frac{V}{2\omega^p_Z(\mbox{\boldmath $k$},\tau)}\right]
\label{ZZ}
\end{equation}
where $p$ denotes the three possible polarizations,
$\omega^p_Z(\mbox{\boldmath $k$},\tau)$ denotes the dispersion relation
for $Z$ and for the vacuum,

\begin{equation}
{\cal Z}^p_{\rm vac}(\mbox{\boldmath $k$},\tau)=
\left[\frac{V}{2\omega^p_Z(\mbox{\boldmath $k$},\tau)}\right]^{1/2}\;
e^{-i\int \omega^p(\mbox{\scriptsize\boldmath $k$},\tau)d\tau}
\end{equation}

Now, using (\ref{asym0}) and taking \mbox{$\omega_Z\simeq k$}
(\ref{ZZ}) becomes,

\begin{equation}
\langle Z^2\rangle_{_V}\simeq -3\int_0^H\frac{k\,dk}{4\pi^2}
\left[\frac{|\Gamma(\nu)|^2}{\pi}
\left(-\frac{2}{k\tau}\right)^{2\nu-1}-1\;\right]
\end{equation}
Then, considering \mbox{$\frac{1}{2}-\nu\simeq(M_Z/H)^2\ll 1$} we get,

\begin{equation}
\langle Z^2\rangle_{_V}\simeq -3H^2\int_0^1\frac{du}{8\pi^2}
\left[\left(\frac{4}{u}\right)^{-(M_Z/H)^2}-1\;\right]\simeq
\frac{3H^2}{8\pi^2}\Big(\frac{M_Z}{H}\Big)^2<H^2
\label{zhart}
\end{equation}
where \mbox{$u\equiv (k\tau)^2$}. Obviously, one arrives at a similar result 
also for the $W$-bosons.

\subsubsection{The electroweak Higgs field}\label{a22}

Similarly for the EW-Higgs field one has,

\begin{equation}
\langle y^2\rangle_{_V}\equiv 
\langle 0\vert y^2\vert 0\rangle
-\langle 0\vert y^2_{\rm vac}\vert 0\rangle
=\frac{1}{V}\int\!\!\frac{d^3k}{(2\pi)^3}\, 
\left[y^{(1)}(k\tau)y^{(2)}(k\tau)
-\frac{V}{2\omega_\psi(\mbox{\boldmath $k$},\tau)}\right]
\label{yy}
\end{equation}
where $\omega_\psi(\mbox{\boldmath $k$},\tau)$ denotes the dispersion
relation for the scalar field and for the vacuum,

\begin{equation}
y_{\rm vac}(\mbox{\boldmath $k$},\tau)=
\left[\frac{V}{2\omega_\psi(\mbox{\boldmath $k$},\tau)}\right]^{1/2}\;
e^{-i\int \omega_\psi(\mbox{\scriptsize\boldmath $k$},\tau)d\tau}
\end{equation}

Using then (\ref{yasym0}) and also \mbox{$\omega_\psi\simeq k$}, we obtain,

\begin{equation}
\langle y^2\rangle_{_V}\simeq \int_0^H\frac{k\,dk}{4\pi^2}
\left[\frac{|\Gamma(\nu_*)|^2}{\pi}
\left(-\frac{2}{k\tau}\right)^{2\nu_*-1}-1\;\right]
\end{equation}

With \mbox{$\nu_*\simeq\frac{3}{2}$} the above gives,

\begin{equation}
\langle y^2\rangle_{_V}\simeq a^2H^2\int_0^1\frac{du}{8\pi^2}
\left(\frac{1}{u}-1\;\right)\simeq
\frac{a^2H^2}{4\pi^2}\ln\left|\frac{\tau_i}{\tau}\right|
\end{equation}
where the infrared cutoff corresponds to the onset of inflation and, for 
superhorizon scales, \mbox{$k|\tau|\sim k/H\ll 1$}. The above directly 
results in (\ref{ylnt}).\footnote{Here we used that 
\mbox{$k=aH=-1/\tau\Rightarrow\ln|k_1/k_2|=\ln|\tau_2/\tau_1|$}.}

Comparing the backreaction terms in view of (\ref{zhart}) and (\ref{ylnt})
we see that the $3\lambda_*\langle\Psi^\dag\Psi\rangle_{_V}$ term is indeed the
dominant when \mbox{$\lambda_*\geq g^4/2\pi^2\sim 10^{-3}$}.

\section{The rms value of the hypermagnetic field}\label{appb}

Here we will calculate the rms value of the hypermagnetic field 
$B_{\rm rms}^Y(\ell)$ as a function of scale at the end of inflation.

The definition of the hypermagnetic field implies,

\begin{equation}
{\cal B}^Y_i(\mbox{\boldmath $k$};t)
= i\epsilon^{ijl}k_j {\cal Y}_l=i\epsilon^{ijl}k_j {\cal Y}_l^\perp
\label{Bi}
\end{equation}
where we used, \mbox{$\epsilon^{ijl}k_j {\cal Y}_l^\parallel=0$} and 

\begin{eqnarray}
{\cal B}^Y_i(\mbox{\boldmath $k$},t)=\int d^3\!x\,
e^{-i\mbox{\scriptsize\boldmath $k\!\cdot\! x$}} 
B^Y_i(\mbox{\boldmath $x$},t) & 
\hspace{1cm} &
B^Y_i(\mbox{\boldmath $x$},t)=\int\frac{d^3 k}{(2\pi)^3}\,
e^{i\mbox{\scriptsize\boldmath $k\!\cdot\! x$}} 
{\cal B}^Y_i(\mbox{\boldmath $k$},t)
\label{Bidef}\\
 & & \nonumber\\
{\cal Y}_i(\mbox{\boldmath $k$},t)=\int d^3\!x\,
e^{-i\mbox{\scriptsize\boldmath $k\!\cdot\! x$}} 
Y_i(\mbox{\boldmath $x$},t) & 
\hspace{1cm} &
Y_i(\mbox{\boldmath $x$},t)=\int\frac{d^3 k}{(2\pi)^3}\,
e^{i\mbox{\scriptsize\boldmath $k\!\cdot\! x$}} 
{\cal Y}_i(\mbox{\boldmath $k$},t)
\label{Yidef}
\end{eqnarray}

We now define,

\begin{equation}
(B_{\rm rms}^Y)^2=\langle \bar{B}_i^Y\!(\ell,\mbox{\boldmath $x$})\,
\bar{B}_i^Y\!(\ell,\mbox{\boldmath $x$})\rangle\equiv
\frac{1}{V}\int d^3\!x\,
\bar{B}_i^Y(\ell,\mbox{\boldmath $x$})
\bar{B}_i^Y(\ell,\mbox{\boldmath $x$})
\end{equation}
where

\begin{equation}
\bar{B}_i^Y\!(\ell,\mbox{\boldmath $x$})
=\frac{1}{V}
\int_{_{\!|\mbox{\tiny\boldmath $x$}-\mbox{\tiny\boldmath $x$}'|\leq\ell}}
\hspace{-0.8cm}d^3\!x'\;B_i^Y\!(\mbox{\boldmath $x$}')
\end{equation}

Thus, in view of the above we have, 

\begin{equation}
(B_{\rm rms}^Y)^2=\frac{1}{V^3}\int d^3\!x
\int_{_{\!|\mbox{\tiny\boldmath $x$}-\mbox{\tiny\boldmath $y$}|\leq\ell}}
\hspace{-0.6cm}d^3\!y
\int_{_{\!|\mbox{\tiny\boldmath $x$}-\mbox{\tiny\boldmath $z$}|\leq\ell}}
\hspace{-0.6cm}d^3\!z
\int\frac{d^3k\,d^3k'}{(2\pi)^6}
\exp[i(\mbox{\boldmath $k\cdot y$}+\mbox{\boldmath $k$}'
\mbox{\boldmath $\cdot\,z$})]\,
\mbox{\boldmath $\cal B$}^Y(\mbox{\boldmath $k$})\mbox{\boldmath $\cdot$}
\mbox{\boldmath $\cal B$}^Y(\mbox{\boldmath $k$}')
\label{int}
\end{equation}

Now, since 
\mbox{{\boldmath $\cal B$}$^Y$= 
$i${\boldmath $k$}$\times${\boldmath $\cal Y$}$^\perp$({\boldmath $k$})} 
we have, 

\begin{equation}
\mbox{\boldmath $\cal B$}^Y(\mbox{\boldmath $k$})\mbox{\boldmath $\cdot$}
\mbox{\boldmath $\cal B$}^Y(\mbox{\boldmath $k$}')=
[\,\mbox{\boldmath $k\!\cdot{\cal Y}$}^\perp(\mbox{\boldmath $k$}')]
[\,\mbox{\boldmath $k$}'
\mbox{\boldmath $\cdot{\cal Y}$}^\perp(\mbox{\boldmath $k$})]
-(\mbox{\boldmath $k\cdot k$}')
[\mbox{\boldmath $\cal Y$}^\perp(\mbox{\boldmath $k$})
\mbox{\boldmath $\cdot{\cal Y}$}^\perp(\mbox{\boldmath $k$}')]
\end{equation}

Using this and substituting 
\mbox{{\boldmath $K$} $\equiv\frac{1}{2}$
({\boldmath $k$} + {\boldmath $k$}$'$)},
\mbox{{\boldmath $\kappa$} 
$\equiv\frac{1}{2}$({\boldmath $k$} $-$ {\boldmath $k$}$'$)} and
\mbox{{\boldmath $r$}$_1\equiv$ {\boldmath $y$} $-$ {\boldmath $x$}},
\mbox{{\boldmath $r$}$_2\equiv$ {\boldmath $z$} $-$ {\boldmath $x$}}
we can recast (\ref{int}) as,

\begin{eqnarray}
(B_{\rm rms}^Y)^2\hspace{-0.2cm} & = & 
\hspace{-0.2cm}\frac{8}{V^3}\int d^3\!x
\int_{_{\!|\mbox{\tiny\boldmath $r$}_{_1}|\leq\ell}}
\hspace{-0.6cm}d^3\!r_1
\int_{_{\!|\mbox{\tiny\boldmath $r$}_{_2}|\leq\ell}}
\hspace{-0.6cm}d^3\!r_2
\int\frac{d^3K\,d^3\kappa}{(2\pi)^6}
\exp\big\{i[\mbox{\boldmath $K\cdot$}
(\mbox{\boldmath $r$}_1+\mbox{\boldmath $r$}_2+
2\mbox{\boldmath $x$})+\mbox{\boldmath $\kappa\cdot$}
(\mbox{\boldmath $r$}_1-\mbox{\boldmath $r$}_2)]\big\}\times
\label{int3}\\
 & & \nonumber\\
& \hspace{-3.5cm}\times & \hspace{-2cm}
\Big\{
\mbox{\boldmath $K\cdot$}
[\mbox{\boldmath $\cal Y$}^\perp(\mbox{\boldmath $K$+$\kappa$})+
\mbox{\boldmath $\cal Y$}^\perp(\mbox{\boldmath $K$-$\kappa$})]
-\mbox{\boldmath $\kappa\cdot$}
[\mbox{\boldmath $\cal Y$}^\perp(\mbox{\boldmath $K$+$\kappa$})-
\mbox{\boldmath $\cal Y$}^\perp(\mbox{\boldmath $K$-$\kappa$})]+
(\kappa^2\!-\!K^2)
[\mbox{\boldmath $\cal Y$}^\perp(\mbox{\boldmath $K$+$\kappa$})
\mbox{\boldmath $\cdot{\cal Y}$}^\perp(\mbox{\boldmath $K$-$\kappa$})]\Big\}
\nonumber
\end{eqnarray}
where \mbox{$K=|${\boldmath $K$}$|$} and 
\mbox{$\kappa=|${\boldmath $\kappa$}$|$}.
Since, \mbox{$\int d^3\!x\int\frac{d^3K}{(2\pi)^3}
e^{2i\mbox{\scriptsize\boldmath $K\!\cdot\!x$}}=
\frac{1}{8}\,\delta$({\boldmath $K$})} the above becomes, 

\begin{equation}
(B_{\rm rms}^Y)^2=\frac{4}{V^3}
\int_{_{\!|\mbox{\tiny\boldmath $r$}_{_1}|\leq\ell}}
\hspace{-0.6cm}d^3\!r_1
\int_{_{\!|\mbox{\tiny\boldmath $r$}_{_2}|\leq\ell}}
\hspace{-0.6cm}d^3\!r_2
\int\frac{d^3k}{(2\pi)^3}
\;e^{i\mbox{\scriptsize\boldmath $k\cdot$}
(\mbox{\scriptsize\boldmath $r$}_1-\mbox{\scriptsize\boldmath $r$}_2)}
k^2|{\cal Y}(k)|^2
\label{int4}
\end{equation}
where we put \mbox{{\boldmath $\kappa$} $\rightarrow$ {\boldmath $k$}} 
and\footnote{Because the coefficients for {\boldmath $\cal Z$}$^\perp$
in (\ref{Z}) do not depend on direction, it is reasonable to assume that,
on average, the hypermagnetic field depends only on the magnitude 
of the momentum, so that
\mbox{{\boldmath ${\cal Y}$}$^\perp(${\boldmath $k$}$)=$
{\boldmath ${\cal Y}$}$^\perp(k)$}, where \mbox{$k=|${\boldmath $k$}$|$}.}
\mbox{{\boldmath $\cal Y$}$^\perp$({\boldmath $k$})
{\boldmath $\!\!\cdot{\cal Y}$}$^\perp$(-{\boldmath $k$}) 
$\simeq 4|{\cal Y}(k)|^2$} (because the transverse component has two 
polarizations) while also using,
\mbox{{\boldmath $k\!\cdot\!{\cal Y}$}$^\perp$({\boldmath $k$})=
{\boldmath $k\!\cdot\!{\cal Y}$}$^\perp$(-{\boldmath $k$}) = 0}. 

It is easy to show that,

\begin{equation}
\int_{_{\!|\mbox{\tiny\boldmath $r$}|\leq\ell}}\hspace{-0.4cm}d^3\!r
\;e^{\pm i\mbox{\scriptsize\boldmath $k\cdot r$}}=
\int_0^\ell 2\pi r^2dr\int_{-1}^1 d(\cos\vartheta)\,e^{\pm ikr\cos\vartheta}=
\frac{4\pi}{k^2}\Big[\frac{1}{k}\sin(k\ell)-\ell\cos(k\ell)\Big]
\end{equation}
where \mbox{$\cos\vartheta\equiv$ ({\boldmath $k\cdot r$})$/kr$}
and \mbox{$r=|${\boldmath $r$}$|$}.
Using the above into (\ref{int4}) we obtain equation (\ref{Brms0}),

\begin{eqnarray}
[B_{\rm rms}^Y(\ell)]^2 & = & \frac{18(\sin\theta_W)^2}{\pi^2\ell^6}
\int_0^{+\infty} dk\Big[\frac{1}{k}\sin(k\ell)-\ell\cos(k\ell)\Big]^2
\frac{|{\cal Z}(k)|^2}{V}
\nonumber
\end{eqnarray}
\vspace{-1.2cm}
\begin{flushright}
(\ref{Brms0})
\end{flushright}
where, from (\ref{Yrms}), 
\mbox{{\boldmath ${\cal Y}$}$^\perp\simeq
-\sin\theta_W${\boldmath ${\cal Z}$}$^\perp$} and
we took \mbox{$V=\frac{4}{3}\pi\ell^3$}. Since we expect that the 
contribution of the modes with \mbox{$k>1/\ell$} is negligible we may 
consider only the superhorizon spectrum of ${\cal Z}^\perp(k)$, which is 
the one given by (\ref{Zk}). Inserting the latter into (\ref{Brms0}) and
after some algebra we find,

\begin{equation}
[B_{\rm rms}^Y(\ell)]^2=\frac{9}{2\pi^2\ell^2}
\frac{(\sin\theta_W)^2H^2}{(2\ell H)^{^{2(M_Z/H)^2}}}\Big(\frac{M_Z}{H}\Big)^4
\int_0^{+\infty} \frac{dw}{w}\Big[\frac{\sin w}{w^2}-\frac{\cos w}{w}\Big]^2
\label{Brms2}
\end{equation}
where \mbox{$w\equiv k\ell$}. Using spherical Bessel's functions the 
integral $I$ in the above evaluates to, \mbox{$I=1/4$}. Therefore, we
obtain,

\begin{equation}
B_{\rm rms}^Y(\ell)=
\frac{3\sqrt{2}\,\sin\theta_W}{2\ell(2\ell H)^{^{(M_Z/H)^2}}}
\Big(\frac{H}{2\pi}\Big)
\Big(\frac{M_Z}{H}\Big)^2
\propto \ell^{-1-(M_Z/H)^2}
\label{Brms3}
\end{equation}
In fact, one obtains, \mbox{$B_{\rm rms}^Y(\ell)\propto \ell^{\nu-3/2}$}
which, in the limit when \mbox{$M_Z\rightarrow H/2$} 
($\nu\rightarrow 0$) corresponds to the thermal spectrum of massless 
particles, \mbox{$B\propto\ell^{-3/2}$}. However, we actually have 
\mbox{$(M_Z/H)^2\ll 1$} so that, the above becomes (\ref{Brms}).

In a similar way one can obtain the rms value of the $Z$-boson field over
superhorizon scales. One finds,

\begin{equation}
Z_{\rm rms}(\ell)=
\frac{\sqrt{2\,\Gamma[(M_Z/H)^2]}}{
8(2\ell H)^{^{(M_Z/H)^2}}}
\Big(\frac{H}{2\pi}\Big)
\Big(\frac{M_Z}{H}\Big)^2
\propto\ell^{-(M_Z/H)^2}
\end{equation}

We see that the spectrum is almost scale invariant. Note, that the value of 
the Gamma function is very large for \mbox{$(M_Z/H)^2\ll 1$}. Physically, this
is because the scale invariance of the spectrum results in all the modes with 
\mbox{$k<1/\ell$} contributing substantially to the rms value.

\end{document}